\documentclass[prd,amsmath,amssymb,twocolumn,tightenlines,floatfix,notitlepage,preprintnumbers,superscriptaddress,nofootinbib]{revtex4}

\usepackage{graphicx}
\usepackage{xspace}
\usepackage[small]{subfigure}
\usepackage{color}

\newcommand{\nn}{\nonumber} 
\newcommand{\bn}{{\bar n}}
\newcommand{\mcdot}{\!\cdot\!}

\newcommand{\be}{\begin{equation}}
\newcommand{\ee}{\end{equation}}

\newcommand{\vect}[1]{\mathbf{#1}}
\newcommand{\abs}[1]{\left\lvert #1\right\rvert}
\newcommand{\bra}[1]{\left\langle #1\right\rvert}
\newcommand{\ket}[1]{\left\lvert #1\right\rangle}

\newcommand{\minus}{\!-\!}
\newcommand{\plus}{\!+\!}

\newcommand{\as}{\alpha_s}
\newcommand{\MSbar}{\overline{\text{MS}}}

\newcommand{\kt}{${\rm k_T}$\xspace}

\newcommand{\cO}{\mathcal{O}}
\newcommand{\cM}{\mathcal{M}}
\newcommand{\cC}{\mathcal{C}}
\newcommand{\cI}{\mathcal{I}}
\newcommand{\cT}{\mathcal{T}}
\newcommand{\cG}{\mathcal{G}}
\newcommand{\cQ}{\mathcal{Q}}
\newcommand{\cH}{\mathcal{H}}

\newcommand{\cA}{\mathcal{A}}

\newcommand{\ca}[1]{\mathcal{#1}}

\newcommand{\e}{\epsilon}

\newcommand{\Scum}{{\cal S}^c}

\newcommand{\eq}[1]{Eq.~\eqref{#1}}
\newcommand{\eqs}[2]{Eqs.~\eqref{#1} and \eqref{#2}}

\renewcommand{\sec}[1]{Sec.~\ref{#1}}

\newcommand{\appx}[1]{App.~\ref{#1}}
\newcommand{\fig}[1]{Fig.~\ref{#1}}
\newcommand{\figs}[2]{Figs.~\ref{#1} and \ref{#2}}

\allowdisplaybreaks[3]

\DeclareMathOperator{\Li}{Li}

\begin{document}

\preprint{MIT-CTP 4302, INT-PUB-11-043}

\title{Double Non-Global Logarithms In-N-Out of Jets}
\author{Andrew Hornig}
\affiliation{University of Washington, Seattle, WA  98195-1560, USA}
\author{Christopher Lee}
\affiliation{Center for Theoretical Physics,  Massachusetts Institute of Technology, Cambridge, MA, 02139, USA}
\author{Jonathan R. Walsh}
\affiliation{Theoretical Physics Group, Ernest Orlando Lawrence Berkeley National Laboratory, 
and Center for Theoretical Physics, University of California, Berkeley, CA 94720, USA }
\author{Saba Zuberi}
\affiliation{Theoretical Physics Group, Ernest Orlando Lawrence Berkeley National Laboratory, 
and Center for Theoretical Physics, University of California, Berkeley, CA 94720, USA }
\date{\today}
\preprint{}

\begin{abstract}
We derive the leading non-global logarithms (NGLs) of ratios of jet masses $m_{1,2}$ and a jet energy veto $\Lambda$ due to soft gluons splitting into regions in and out of jets. Such NGLs appear in any exclusive jet cross section with multiple jet measurements or with a veto imposed on additional jets. Here, we consider back-to-back jets of radius $R$ produced in $e^+e^-$ collisions, found with a cone or recombination algorithm. The leading NGLs are of the form $\as^2 \ln^2(\Lambda/m_{1,2})$ or $\as^2\ln^2(m_1/m_2)$. Their coefficients depend both on the algorithm and on $R$. We consider cone, \kt, anti-\kt, and Cambridge-Aachen algorithms. In addition to determining the full algorithmic and $R$ dependence of the leading NGLs, we derive new relations among their coefficients.  We also derive to all orders in $\as$ a factorized form for the soft function $S(k_L,k_R,\Lambda)$  in the cross section $\sigma(m_1,m_2,\Lambda)$ in which dependence on each of the global logs of $\mu/k_L$, $\mu/k_R$ and $\mu/\Lambda$ determined by the renormalization group are separated from one another and from the non-global logs. The same kind of soft function, its associated non-global structure, and the algorithmic dependence we derive here will also arise in exclusive jet cross sections at hadron colliders, and must be understood and brought under control to achieve precise theoretical predictions.
\end{abstract}

\maketitle

\section{Introduction}

Observables of varying exclusivity can be used to probe the jetlike structure of final states in high energy collisions \cite{Dasgupta:2003iq,Banfi:2004nk} or even the substructure of the jets themselves \cite{Abdesselam:2010pt}. While more exclusivity reveals more information about structure, it also introduces dependence on additional scales, ratios of which induces  potentially large logarithms in perturbative expansions. 

Non-global observables  \cite{Dasgupta:2001sh,Dasgupta:2002bw} are those for which soft radiation in sharply divided regions of phase space are probed with different measures. For example, measuring separate masses $m_{1,2}$ of two back-to-back jets produced in $e^+e^-$ collisions generates non-global logarithms (NGLs) of $m_1/m_2$. Measuring the total invariant mass of a two-jet-like final state, $m^2 = m_1^2 + m_2^2$, however, generates no NGLs since the soft radiation everywhere is probed ``globally'' with a single scale $m$. Global logs like those of $m/Q$ in such a cross section can be resummed using well-known methods (e.g. \cite{Catani:1992ua}). The most powerful of these are based on the renormalization group (RG) evolution of hard, jet and soft functions in a factorization theorem for the global observable in perturbative QCD \cite{Sterman:1995fz,Contopanagos:1996nh} or soft-collinear effective theory (SCET) \cite{Bauer:2000ew,Bauer:2000yr,Bauer:2001ct,Bauer:2001yt,Bauer:2002nz}. 

To be more precise, for the example of the hemisphere dijet mass distribution, a factorization theorem for the distribution in $m_{1,2}^2$ takes the form  \cite{Fleming:2007qr,Fleming:2007xt}
\be
\label{dijetcsfactorized}
\begin{split}
\sigma(m_1^2,m_2^2) &= \sigma_0 H(Q,\mu) \int \! dk_L dk_R
J_\bn(m_1^2 \minus Qk_L,\mu)  \\
&\quad\times J_n (m_2^2 \minus Qk_R,\mu) S(k_L,k_R,\mu) + \cdots \,,
\end{split}
\ee
where $k_{L,R}$ measure the small light-cone components $\bn\cdot k,n\cdot k$ of the total momenta in each hemisphere, where $n,\bn = (1,\pm \vect{\hat z})$ with $+\vect{\hat z}$ along the jet axis in the right (R) hemisphere. The soft function $S$ can be expressed as a convolution of perturbative and nonperturbative  pieces \cite{Hoang:2007vb}, but we only consider its perturbative component here. The hard function $H$ depends only on logs of $\mu/Q$ and the jet functions $J_{1,2}$ depend on logs of $\mu/m_{1,2}$, but the soft function depends on logs of $\mu/k_L$, $\mu/k_R$, and $k_L/k_R$. The $\mu$-dependent ``global'' logs can be resummed by RG evolution, but the logs of $k_L/k_R$ are non-global and not resummed by the ordinary RG. In other words, logs of the ratio of any single soft scale to the hard scale or either jet scale can be resummed by running between those scales. But logs of ratios of soft scales among themselves cannot be resummed by using a framework that only contains one soft mode and thus one soft scale to or from which RG evolution can be performed.

Another case in which multiple soft scales appear is when two jets $i,j$ in a multijet event come close together.  This introduces the soft scales $m_i^2 / m_{ij}$ and $m_j^2 / m_{ij}$, associated with the ``fat'' dijet, in to the cross section, in addition to the usual soft scales,  $m_i^2/Q$, associated with individual jets.  Logs induced by these additional soft scales  can be summed using the effective theory SCET$_+$, an extension of SCET$_{\rm I}$ that contains a ``collinear-soft'' (csoft) mode \cite{Bauer:2011uc}.  This was the first factorization of a multi-scale soft function that allows for complete resummation of all the resulting large logarithms in the cross section. These logs should be distinguished from non-global logs, which come from making measurements in different regions of phase space and are ratios of soft scales (e.g., $m_i/m_j$).

NGLs were first recognized in \cite{Dasgupta:2001sh} in  $e^+ e^-$ dijet event shape distributions in which the mass of only one hemisphere jet $\rho_R = m_R/Q$ is measured while being inclusive in the other hemisphere, and in a larger class of event shapes in deep inelastic scattering in \cite{Dasgupta:2002dc}. Subsequently, Refs.~\cite{Dasgupta:2002bw,Banfi:2002hw,Appleby:2002ke} studied NGLs of $\Lambda/Q$ in cross sections vetoing radiation with total energy greater than $\Lambda$ in angular regions outside of found jets. Though a hard scale $Q$ appears in these ratios, we found in \cite{Hornig:2011iu} that the NGLs still arise from considering both scales in the ratio to be soft and later taking one of them to $Q$ in an inclusive limit.

In \cite{Hornig:2011iu} we made progress in understanding the origin of NGLs in effective field theory. We considered the factorized dijet invariant mass distribution $\sigma(m_1,m_2)$ in $e^+e^-$ collisions producing back-to-back jets, and calculated to $\cO(\as^2)$, as also in \cite{Kelley:2011ng}, the hemisphere soft function $S(k_L,k_R)$. These calculations clarified the origin of NGLs in an EFT framework as the dependence of a soft function on ratios of multiple soft scales, and revealed new subleading (single)  NGLs and non-logarithmic non-global functions.

These NGLs are organized into a multiplicative factor entering the total cross section, with the leading NGLs taking the generic form
\be \label{eq:NGLform}
S_{\text{NG}}(\mu_1/\mu_2) = 1 -\frac{\as^2}{(2\pi)^2} C_F C_A S_2 \ln^2\frac{\mu_1}{\mu_2} +\cdots \,.
\ee
Here $\mu_{1,2}$ are the scales at which soft radiation is probed in different sharply-divided regions. For the hemisphere mass distribution $\mu_{1,2} = m_{1,2}^2/Q$ and $S_2 =\pi^2/3$. For the $\rho_R$ distribution, $\mu_1 = Q\rho_R$ while $\mu_2=Q$ due to total inclusivity in one hemisphere. The coefficient $S_2$ is a geometric measure of the region into which the two soft gluons contributing to a NGL can go. The fact that it varies with the size of this region is due to the NGL arising from a purely soft divergence of QCD.  Techniques to resum NGLs using numerical fits in the large-$N_C$ limit of QCD were introduced by \cite{Dasgupta:2001sh}, but analytic resummation of NGLs in real-world QCD remains an open problem.

\begin{figure}[t]{
    \includegraphics[width=.88\columnwidth]{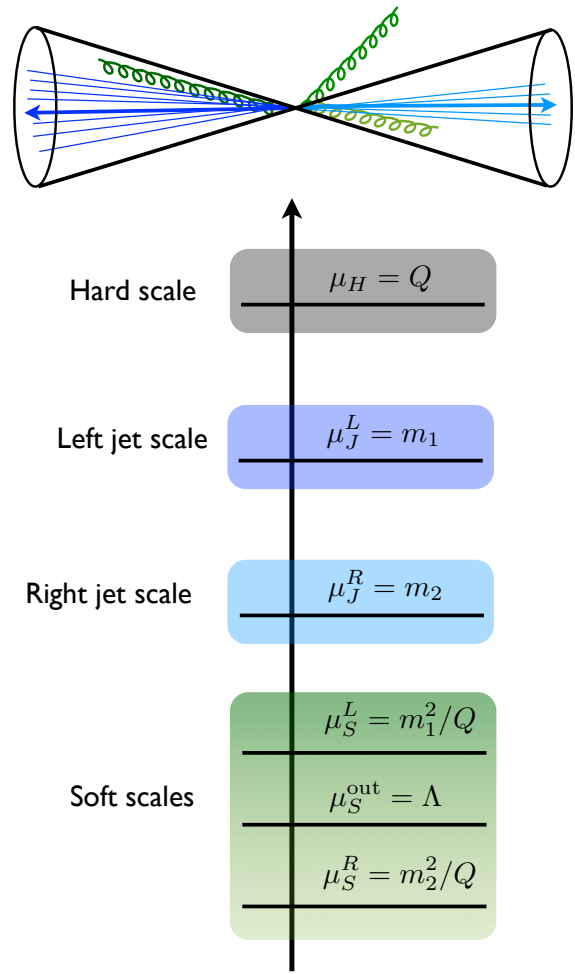}\vspace{-1ex} { \caption[1]{The relevant scales in the exclusive jet mass cross section with an energy veto, $\Lambda$ outside of the jets is shown for a particular choice of the hierarchy $m_2^2 \ll \Lambda Q \ll m_1^2 $ that gives rise to large non-global logs. Our results apply to any choice of $m_{1,2}$ and $\Lambda$ that satisfies $Q \gg m_{1,2} \gg m_{1,2}^2/Q,\,\Lambda$, which maintains the separation between hard, jet and soft scales. }
  \label{fig:scalesFig}} }
\end{figure}

In this work we seek  to extend the intuition gained in \cite{Hornig:2011iu} by studying a more exclusive set of cross sections. We study non-global properties of an exclusive jet cross section $\sigma(m_1,m_2,\Lambda)$, where the invariant masses $m_1$ and $m_2$ of two jets of size $R$ produced in an $e^+e^-$ collision at center-of-mass energy $Q$ are measured, with a veto $\Lambda$ on the energy of additional jets. We consider finding the jets using various algorithms---cone, anti-\kt, Cambridge-Aachen, and \kt \cite{Catani:1991hj,Catani:1993hr,Ellis:1993tq,Dokshitzer:1997in,Salam:2007xv,Cacciari:2008gp}. We will find that NGLs of the ratio of the jet veto and the jet masses $\Lambda/m_{1,2}$ are present, in addition to NGLs of the ratio of masses $m_1/m_2$. We calculate the coefficients only of leading double NGLs $\as^2\ln^2(\mu_1/\mu_2)$ in this paper. The relevant scales for this observable are shown in \fig{fig:scalesFig} for a particular hierarchy of $m_{1,2}$ and $\Lambda$, however our results are valid for any choice such that $Q \gg m_{1,2} \gg m_{1,2}^2/Q,\,\Lambda$.

\begin{figure*}[th]{
    \includegraphics[width=1.0\textwidth]{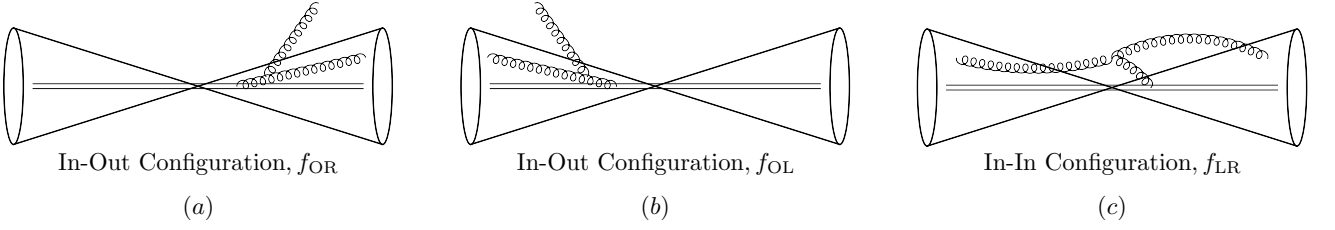}\vspace{-3ex} { \caption[1]{The three basic soft gluon configurations that we use to calculate the leading NGLs.  At $\ca{O}(\alpha_s^2)$, the coefficient $f_{\text{OR}}$ ($f_{\text{OL}}$) receives contributions from one gluon in the right (left) jet and one gluon out of both jets, and the coefficient $f_{\text{LR}}$ receives contributions from one gluon in each jet.}
  \label{fig:inoutconfigs}} }
\end{figure*}

In \cite{Hornig:2011iu}, we discovered that at $\cO(\as^2)$ NGLs of two soft scales $\mu_{1,2}$ can be constructed from separate pieces dependent on the ratio of the factorization scale $\mu$ to one physical scale at a time. Namely, the region of phase space where one of the soft gluons enters the  region sensitive to the scale $\mu_1$ and the other enters the region sensitive to $\mu_2$ generates the double log $\as^2\ln^2\mu^2/(\mu_1\mu_2)$, while the regions where soft gluons enter only region 1 or only region 2 generate   $\as^2\ln^2(\mu/\mu_1)$ and $\as^2\ln^2(\mu/\mu_2)$. In \cite{Hornig:2011iu} we derived from RG invariance of the cross section and IR safety of the soft function that the coefficients of these logs are constrained so that the $\mu$-dependence cancels, but an NGL $\as^2\ln^2(\mu_1/\mu_2)$ is left over. Analogously for $\sigma(m_1,m_2,\Lambda)$, the three soft phase space regions that give rise to the NGLs at $\cO(\as^2)$ are shown in \fig{fig:inoutconfigs}.  Each configuration contributes logarithms of $\mu$ over a single scale, the ``in-out'' regions contributing logs $\as^2\ln^2\mu^2 / (\Lambda \,m_{1,2})$, and the ``in-in'' region contributing logs  $\as^2\ln^2\mu^2 / (m_1m_2)$. These combine with single-region contributions to give NGLs of $\Lambda/m_{1,2}$ with coefficients $f_{\text{OL,OR}}$ and of $m_1/m_2$ with coefficient $f_{\text{LR}}$. These coefficients give the geometric factor  $S_2$ in \eq{eq:NGLform}. IR safety and RG invariance will allow us to derive additional strong relations among these different coefficients.

The division of the NGL into separately $\mu$-dependent pieces opens up the possibility to use the renormalization group to sum NGLs, although this has yet to be carried out explicitly.

It is worth noting that our calculation of the coefficients $f_{\text{OL,OR,LR}}$ of the leading NGLs for a two-jet configuration applies to other measurements of the soft radiation in and out of the jets as well. Different choices of observable change the arguments of the NGLs, but their coefficients are related to the geometry of the configuration given in \fig{fig:inoutconfigs} and will be given by one of $f_{\text{OL,OR,LR}}$.

While the presence of NGLs has previously been recognized in jet mass observables similar to the one we consider here, we derive the  full algorithmic and $R$ dependence of the leading NGLs involving both jet masses and vetoes here for the first time. Refs.~\cite{Dasgupta:2002bw,Banfi:2002hw} derived the NGL of $\Lambda/Q$ in cross sections with a veto region outside jets defined with fixed cones (but with masses unmeasured). Ref.~\cite{Appleby:2002ke} studied (and \cite{Delenda:2006nf} calculated more precisely) the effect of clustering soft gluons with a fixed $R=1.0$ with the \kt\ algorithm, with a variable rapidity gap $\Delta\eta$ describing the size of the region in which radiation is vetoed. Ref.~\cite{Banfi:2010pa} considered a cross section with two cone jets of radius $R\ll 1$, with only one jet's mass $m$ being measured and a veto  $\Lambda$ on radiation outside the jets, and calculated the NGLs of $m/\Lambda$ due to the measured jet and $\Lambda/Q$ due to the inclusive jet. Ref.~\cite{Rubin:2010fc} was the first to study NGLs in an observable probing jet substructure, the mass drop in filtered subjets produced by decays of boosted Higgs bosons \cite{Butterworth:2008iy}.   Our work consolidates and extends many of these results by calculating NGLs of $\Lambda/m_{1,2}$ and $m_1/m_2$ with two measured jets for four different cone and recombination algorithms, as a function of arbitrary jet size $0<R<\pi/2$ \footnote{We do require $R$ to be large enough for the observable we consider to factorize, namely $m_{1,2} \ll Q \tan(R/2)$ \cite{Ellis:2009wj,Ellis:2010rw}.}, and deriving new relations among the coefficients of the different NGLs.

For the \kt algorithm, it was pointed out in \cite{Banfi:2005gj,Delenda:2006nf} and elaborated in \cite{Banfi:2010pa} that the effects of soft gluon clustering also affect the independent emission contributions to the types of observables mentioned above. The factorization theorems and resummed predictions we consider below should be modified to include such effects for clustering algorithms. We leave this outside the scope of our present work, focusing just on how NGLs affect such predictions.

We gain some intuition from our investigation that is not necessarily novel, but is often not appreciated, and that we hope is helpful to clarify. First, the presence of multiple soft scales is enough to induce NGLs, regardless of their ordering. Second, we emphasize that NGLs arise not only from soft gluons splitting right along jet boundaries and entering just inside the respective separate regions, but from the entire angular region of phase space into which the soft gluons can enter. The numerically largest contribution comes from the two gluons being close to each other near the boundary, but the enhancement is not parametrically large. Thus NGLs cannot be avoided simply by erasing dependence on the boundary region in constructing a jet observable, although they can be somewhat reduced. Probing separate regions  with separate soft scales is enough to induce NGLs \cite{Appleby:2002ke}.

While our results here are derived for $e^+ e^-$ collisions, the methods and lessons are directly applicable to exclusive jet cross sections measured at hadron colliders such as the LHC. For instance, distributions in multiple jet masses, jet shapes or event shapes such as $N$-jettiness \cite{Stewart:2010tn,Jouttenus:2011wh} with different values of the measure on the $N$ jets will contain NGLs of $m_i/m_j$ or $\tau_1/\tau_2$. Exclusive jet cross sections defined with explicit vetoes on the $p_T$ of additional jets induce NGLs of $p_T^{\text{cut}}$ over the relevant hard scale. For example, vetoing jets in searches for Higgs to $WW\to \ell\nu \ell\nu$ with zero jets will induce NGLs of $p_T^{\text{cut}}/m_H$. It would thus be wise either to calculate and control these NGLs or to use methods to veto jets that avoid NGLs, such as beam thrust (0-jettiness) based vetoes \cite{Stewart:2010tn,Stewart:2010pd,Berger:2010xi}.

A final consequence of our results regards the claim of the recent Ref.~\cite{Kelley:2011tj} by Kelley, Schwartz, and Zhu (KSZ) that the cross section $\sigma(\rho,\Lambda)$, where $\rho = (m_1^2+m_2^2)/Q^2$, contains no logs of $\Lambda/(Q\rho)$,  and that therefore a factorization of the form $\sigma(\rho,\Lambda) = \sigma_{\text{in}}(\rho)\sigma_{\text{out}}(\Lambda)$ holds to all orders in $\as$, at least in the regime $\Lambda<Q\rho\ll Q R\ll 1$\footnote{We note that this claim appeared in version 1 of \cite{Kelley:2011tj}, and has since been retracted in later versions.}. This conclusion is not consistent with our calculations, which show that NGLs of $\Lambda/(Q\rho)$ are in fact present for any value of this ratio. We verified our prediction for the coefficient of the NGL by comparing to the predictions of the numerical Monte Carlo EVENT2 \cite{Catani:1996jh,Catani:1996vz}.
 
In \sec{sec:softfunction} we review the factorization theorem for the cross section $\sigma(m_1,m_2,\Lambda)$ and derive to all orders in $\as$ a correctly factorized form for the soft function $S(k_L,k_R,\Lambda)$ that appears therein. In \sec{sec:doubleNGLs}, we derive new generic relations among coefficients of the different NGLs appearing in $S(k_L,k_R,\Lambda)$. In \sec{sec:NGL} we derive the leading NGLs appearing in $S(k_L,k_R,\Lambda)$, including the full algorithmic and $R$ dependence. In \sec{sec:comparison} we compare the predictions of $\sigma(\rho,\Lambda)$  with and without the predicted NGL of $Q\rho/\Lambda$ to those of EVENT2   at $\cO(\as^2)$, and confirm the presence and predicted sizes of the NGLs. In \sec{sec:conclusions} we conclude. In two Appendices, we provide the ingredients necessary to construct the global logs in $\sigma(m_1,m_2,\Lambda)$ from the RG, and the Feynman diagram amplitudes necessary to calculate the leading NGLs.

\section{Soft Function for Two Jet Masses and a Veto}
\label{sec:softfunction}

Factorization and resummation of (global) logarithms for exclusive jet cross sections defined with cone or recombination algorithms with an energy veto outside the jets was first performed in \cite{Ellis:2009wj,Ellis:2010rw}. They imply that $\sigma(m_1,m_2,\Lambda)$ factorizes in the form
\begin{align} \label{factorization}
\sigma(m_1,m_2 ,\Lambda) & = \sigma_0 H(Q;\mu) \int dk_L dk_R J_{\bn}(m_1^2 - Qk_L;\mu) \nn \\
&  \times J_n(m_2^2 - Qk_R;\mu) S(k_L,k_R,\Lambda;R;\mu) \,.
\end{align}
$\sigma_0$ is the Born cross section for $e^+e^-\to q\bar q$, $J_{n,\bn}$ are jet functions, and $S$ is the soft function. The scales that appear in this factorization theorem are depicted in \fig{fig:scalesFig}.
We will also consider the cumulant distribution, defined by
\be
\Sigma(\rho_1,\rho_2,\Lambda) \equiv \int_{-\infty}^{Q\rho_1} \!\! dm_1 \int_{-\infty}^{Q\rho_2} \!\! dm_2 \int_{-\infty}^\Lambda \!\! d\Lambda'  \sigma(m_1,m_2,\Lambda')
\ee
which also factorizes in the form \cite{Hornig:2011iu}
\be
\begin{split}
\Sigma(\rho_1,\rho_2,\Lambda) &= \sigma_0  H(Q,\mu)  \int \! dk_L dk_R   J_{\bn}(Q\rho_1 \minus k_L,\mu) \\
&\quad\times \! J_n(Q\rho_2 \minus  k_R,\mu)   \Scum(k_L,k_R,\Lambda;R;\mu)\,,
\end{split}
\ee
where $\Scum$ is the cumulant soft function. 
The jet functions also depend on the jet size $R$ and on the algorithm \cite{Cheung:2009sg,Ellis:2009wj,Ellis:2010rw,Jouttenus:2009ns} but, as shown in these references, in the limit $m_{1,2}\ll Q\tan(R/2)$, the dependence on $R$ is power suppressed, and $J_{n,\bn}$ are the usual inclusive jet functions \cite{Bauer:2003pi,Bosch:2004th,Becher:2006qw}. We will work in this limit in what follows.

The soft function $S(k_L,k_R,\Lambda;R)$ arising in \eq{factorization} was  first defined and calculated to $\cO(\as)$ in \cite{Ellis:2009wj,Ellis:2010rw} and is given by
\begin{align} \label{conesoftdef}
S(k_L,k_R & ,  \Lambda;R) = \frac{1}{N_C}\sum_{X_S}  \abs{\bra{X_S} T[ Y_n Y_\bn^\dag] \ket{0}}^2  \nn \\
&\times  \delta\Bigl(\Lambda \minus \sum_{i\in X}\Theta_{\text{out}}k_i^0\Bigr) \\
&\times \delta\Bigl(k_L \minus \sum_{i \in X} \Theta_{\text{in}}^n \bn \mcdot k_i\Bigr) \delta\Bigl(k_R \minus \sum_{i\in X}\Theta_{\text{in}}^\bn n\mcdot k_i\Bigr)  \,. \nn
\end{align}
The theta functions $\Theta_{\text{in,out}}$ choose those soft particles $i$ that end up inside one of the jets or outside both jets. Their precise form depends on the algorithm. In this definition, the energy outside the jets is fixed to be $\Lambda$, but we can integrate to obtain the cumulant which allows all energies up to $\Lambda$.

Much of the structure of the soft function is determined by consistency of the factorization theorem in \eq{factorization} and the RG evolution of the hard, jet, and soft functions. We will argue its perturbative structure must take the form,
\be
\label{factorizedsoft}
\begin{split}
S(k_L,k_R,\Lambda;\mu) &= [S_{\text{in}}(k_L;\mu)S_{\text{in}}(k_R;\mu)] S_{\text{out}}(\Lambda;\mu) \\
&\quad\otimes S_{\text{NG}}(k_L,k_R,\Lambda)\,,
\end{split}
\ee
where $\otimes$ denotes a convolution of $S_{\text{NG}}$ with $S_{\text{in}}$'s in the variables $k_{L,R}$ and with $S_{\text{out}}$ in the variable $\Lambda$.  The cumulant soft function $\Scum$ behaves similarly. The pieces $S_{\text{in,out}}$ are determined by RG evolution and $S_{\text{NG}}$ is not.  $S_{\text{in,out}}$ depend individually on the scales $k_L$, $k_R$, and $\Lambda$, while the $S_{\text{NG}}$ has non-separable dependence on the ratios $k_L /k_R$ and $k_{L,R}/\Lambda$. 

We can derive the form \eq{factorizedsoft} of $S$ from RG invariance of the cross section $\sigma$, which is $\mu$-independent. Since the hard and jet functions (strictly speaking, its Laplace transform) have anomalous dimensions of the form
\be
\mu \frac{d}{d\mu}\ln F = \Gamma_F \ln\frac{\mu^2}{\mu_F^2} + \gamma_F\,,
\ee
where $\mu_F = Q$ for $F=H$ and $\mu_F= Q(\rho_{1,2})^{1/2}$ for $F=  J_{1,2}$, the soft function must have an anomalous dimension
\be
\label{softanomdim}
\mu\frac{d}{d\mu}\ln \Scum = \Gamma_S \ln\frac{\mu^2}{k_L^2}  + \Gamma_S \ln\frac{\mu^2}{k_R^2} + \gamma_S\,,
\ee
where $\Gamma_S = -\Gamma_H/2 - \Gamma_J$ and $\gamma_S = -\gamma_H - 2\gamma_J$. Notably, $\Gamma_S,\gamma_S$ are independent of $R$, and the scale $\Lambda$ cannot appear in \eq{softanomdim} since the hard and jet functions know nothing about $\Lambda$. The pieces $\Gamma_{H,J,S}$ are proportional to the cusp anomalous dimension $\Gamma_{\text{cusp}}$ to all orders in $\as$.

In \cite{Ellis:2009wj,Ellis:2010rw}, we calculated contributions to the soft anomalous dimension at $\cO(\as)$ into pieces coming from a gluon inside a jet or outside the jets, finding $d(\ln\Scum)/d(\ln\mu) = \gamma_L + \gamma_R + \gamma_{\text{out}}$, where
\be
\label{gammaLR}
\gamma_{L,R} = \Gamma_S \ln\Bigl(\frac{\mu \tan\frac{R}{2}}{k_{L,R}}\Bigr)^2 + \gamma_{\text{in}}\,,
\ee
with $\Gamma_S$ the same as in \eq{softanomdim} and $\gamma_{\text{in}}=0$ at $\cO(\as)$. Thus,
\be
\label{gammaout}
\gamma_{\text{out}} = -2\Gamma_S\ln \tan^2\frac{R}{2} + 2\gamma_S - 2\gamma_{\text{in}}\,,
\ee
which was verified by direct calculation to $\cO(\as)$ in \cite{Ellis:2009wj,Ellis:2010rw}. (Note $\gamma_S =0$ also at $\cO(\as)$.)

The soft anomalous dimension can always be split up additively into pieces $\gamma_{L,R},\gamma_{\text{out}}$ of the form in \eqs{gammaLR}{gammaout} to all orders in $\as$. Finding operator or phase space definitions for soft functions which have these as their anomalous dimensions to all orders in $\as$ is another question. We found such definitions which work to $\cO(\as)$ in \cite{Ellis:2009wj,Ellis:2010rw}. Those definitions suggest that the pieces $\gamma_{L,R}$ arise from the UV poles in soft function diagrams with all gluons inside the $L,R$ jets, and $\gamma_{\text{out}}$ from the UV poles in diagrams with all gluons outside the jets. We will call the sum of all such diagrams $S_{L,R}$ and $S_{\text{out}}$, respectively. Here, we only need these sums to be defined as those which have \eqs{gammaLR}{gammaout} as their anomalous dimensions.  It would take further work to show that their operator definitions in \cite{Ellis:2009wj,Ellis:2010rw} have precisely these anomalous dimensions to all orders, but these definitions are not necessary to derive the generic form \eq{factorizedsoft} of the perturbative soft function.

The anomalous dimensions in \eqs{gammaLR}{gammaout} imply that the full soft function $\Scum(k_L,k_R,\Lambda;R;\mu)$  evolves under the RG according to
\be
\label{Sevolved}
\begin{split}
\Scum(k_L,k_R,\Lambda;R;\mu) &= \Scum(k_L,k_R,\Lambda;R;\mu_0)U_{\text{out}}(R;\mu,\mu_0) \\
&\otimes [U_L(k_L;R;\mu,\mu_0)U_R(k_R;R;\mu,\mu_0)]\,,
\end{split}
\ee
where the evolution kernels $U_F = \exp[K_F(\mu,\mu_0) + \omega_F(\mu,\mu_0)\ln(\mu/\mu_F)]$, with $K_F,\omega_F$ defined in terms of $\Gamma_F,\gamma_F$ in \eq{Komegadefs}. 

Splitting up the evolution kernels in \eq{Sevolved} cleverly, we obtain
\begin{align}
\label{Usplit}
&\Scum(k_L,k_R,\Lambda;R;\mu) = \Scum(k_L,k_R,\Lambda;R;\mu_0)\\
&\  \otimes  U_{\text{out}}(R;\mu,2\Lambda) [U_L(k_L;R;\mu,k_L)U_R(k_R;R;\mu,k_R)] \nn \\
&\ \otimes U_{\text{out}}(R;2\Lambda,\mu_0)  [U_L(k_L;R;k_L,\mu_0)U_R(k_R;R;k_R,\mu_0)]\,. \nn
\end{align}
The last two factors in brackets, finally, can be further split into
\be
\label{ULRsplit}
\begin{split}
&U_L(k_L;R;k_L,2\Lambda)\otimes U_L(k_L;R;2\Lambda,\mu_0) \\
&\otimes U_R(k_R;R;k_R,2\Lambda)\otimes U_R(k_R;R;2\Lambda,\mu_0)\,.
\end{split}
\ee
Now all the $\mu_0$-dependent factors in \eqs{Usplit}{ULRsplit} simply evolve $\Scum(k_L,k_R,\Lambda;R;\mu_0)$ in \eq{Usplit} from $\mu_0$ to $2\Lambda$, so no $\mu_0$ dependence actually remains. Then the soft function $\Scum$ (and therefore $S$) automatically takes the factorized form of \eq{factorizedsoft}, with
\begin{subequations}
\begin{align}
 \Scum_{\text{in}}(k_{L},\mu)  &\equiv U_{L}(k_{L};R;\mu,k_{L}) \\
 \Scum_{\text{in}}(k_{R},\mu)  &\equiv U_{R}(k_{R};R;\mu,k_{R}) \\
\Scum_{\text{out}}(\Lambda,\mu) &= U_{\text{out}}(R;\mu,2\Lambda)\,,
\end{align}
\end{subequations}
and
\begin{align}
\label{SNGdef}
\Scum_{\text{NG}}(k_L,k_R,\Lambda) &\equiv  \Scum(k_L,k_R,\Lambda;R,2\Lambda) \\
&\times U_L(k_L;R;k_L,2\Lambda)U_R(k_R;R;k_R,2\Lambda)\,. \nn
\end{align}
All logs in $\Scum$ associated with the anomalous dimensions  are now removed by evaluating $\Scum$ at $\mu=2\Lambda$ in the first line and evolving between $2\Lambda$ and $k_{L,R}$ with the $U_{L,R}$ kernels  in the second line. The anomalous dimension of $S$ in \eq{factorizedsoft} then comes entirely from $S_{\text{in,out}}$, whose forms are now completely determined by the form of $\gamma_{L,R,\text{out}}$ and the RG. The remainder $S_{\text{NG}}$ contains the dependence on the dimensionless ratios $k_L/k_R$ and $\Lambda/k_{L,R}$ in $S$ not constrained by the RG.

It is notable that the form \eq{factorizedsoft} follows just from the form of the soft anomalous dimension \eq{softanomdim} that follows from RG invariance. It would take further work to give operator or phase space definitions of $S_{\text{in,out}}$ that have precisely the anomalous dimensions \eqs{gammaLR}{gammaout} to all orders in $\as$, but the generic form \eq{factorizedsoft} does not depend on such a construction.

Below we will give phase space definitions for $S_{\text{in,out,NG}}$ that work to at least $\cO(\as^2)$. We give results for the $\cO(\as^2)$ terms in $S_{\text{in,out}}$ predicted by RG evolution in \appx{appx:NGL}, and calculate the leading NGLs in $S_{\text{NG}}$ in \sec{sec:NGL}.

\section{Structure of the Non-Global Double Logs}
\label{sec:doubleNGLs}

While the renormalization group does not determine the NGLs in the soft function in \eq{factorization}, constraints from RG invariance can still be used to simplify the method to calculate them \cite{Hornig:2011iu}. Here we use these constraints to derive relationships among the different ``in-in'' and ``in-out'' NGLs in the observables that we consider.

To study the relationships between NGLs, we take the two jet radii to be different, $R_L$ and $R_R$ for the left and right jets respectively.  While this can lead to ambiguities about which $R$ to use in recombination metrics, it is sensible in the $\cO(\as^2)$ calculations we carry out as long as $R_{L,R}$ are sufficiently small, e.g. $R_{L,R}<\pi/3$ (or for any $R_{L,R}$ using fixed cones). In this section we consider different $R$ values to constrain the coefficients of the different possible NGLs and in the next section calculate explicitly the $R_{L,R}$ dependence of the NGLs for the anti-\kt algorithm.  For C/A and \kt, we will calculate for $R_L = R_R = R$. Most of the logic and results in this section actually go through for equal $R$'s as well. We rely on distinct $R_{L,R}$ only at the very end of this section.

The dijet events we consider have three kinds of double non-local globs, each a ratio of scales in the soft sector:
\begin{subequations}
 \label{doubleNGLs}
\begin{align}
\label{OLNGL}
&-\left(\frac{\as}{2\pi}\right)^2 C_F C_A \, f_{\text{OL}} (R_L, R_R) \ln^2 \frac{k_L}{2\Lambda\tan R_L/2} \,,  \\
\label{ORNGL}
&-\left(\frac{\as}{2\pi}\right)^2 C_F C_A \, f_{\text{OR}} (R_R, R_L) \ln^2 \frac{k_R}{2\Lambda\tan R_R/2} \,,  \\
\label{LRNGL}
&-\left(\frac{\as}{2\pi}\right)^2 C_F C_A \, f_{\text{LR}} (R_L, R_R) \ln^2 \frac{k_L \tan R_R/2}{k_R \tan R_L/2} \,.
\end{align}
\end{subequations}
The coefficients $f_{\text{OL}}$, $f_{\text{OR}}$, and $f_{\text{LR}}$ depend on the jet algorithm.  The ratios are between the scales (identified in \cite{Ellis:2009wj,Ellis:2010rw}) of soft gluons inside jets $\mu_S^{L,R} = k_{L,R}/\tan(R_{L,R}/2)$ and the scale $\mu_S^\Lambda = 2\Lambda$ of soft gluons outside jets cutoff by energy $\Lambda$. Contributions to these logs come from three regions of phase space, which we label $L$ for the left jet, $R$ for the right jet, and $O$ for the out-of-jet region.  These regions are shown in \fig{fig:inoutconfigs}. Each  NGL gets a contribution from a pair of these regions, which set the scales in the log.  If we consider two regions $A$ and $B$, then the phase space contributions can be divided as:
\be
\label{PSsplit}
\mathcal{M}(\{k_i\}) = \mathcal{M}_{\text{A}}(\{k_i\}) + \mathcal{M}_{\text{B}}(\{k_i\}) +  \mathcal{M}_{\text{AB}}(\{k_i\})\,,
\ee
where each $\cM$ imposes a set of measurements on the partons in the final state with momenta $k_i$ (cf. \eq{measurements}).
The first term comes from gluons only in region $A$, the second from gluons only in region $B$, and the third from at least one gluon in both regions.

RG evolution constrains the relative contributions from each of the terms in \eq{PSsplit}.  NGLs are independent of the renormalization scale $\mu$, but each contribution from \eq{PSsplit} will have $\mu$-dependence.  For instance, the $f_{\text{OR}}$ NGL comes from the sum:
\begin{align} \label{doublelogexp}
&-\left(\frac{\as}{2\pi}\right)^2 C_F C_A \, f_{\text{OR}} (R_R, R_L) \\
& \quad \times \left[2 \ln^2 \frac{\mu \tan R_R/2}{k_R} + 2\ln^2 \frac{\mu}{2\Lambda} - \ln^2\frac{\mu^2 \tan R_R/2}{2\Lambda k_R} \right] \,. \nn
\end{align}
The last term is especially notable: it only contains contributions with two soft gluons in the final state that live in separate regions (one in $R$, one in $O$), and it is the only term of the three that depends on multiple scales.  These contributions are simpler to compute than the other terms, with the added benefit that there are no global terms with the same color and log structure.  These mixed-scale terms alone determine the coefficient of the NGLs, as the others are fixed by RG invariance.  This  feature was used in \cite{Hornig:2011iu} to determine the complete set of non-global terms in the hemisphere dijet soft function.

Let us consider the bare contribution to the soft function from the last term in \eq{doublelogexp}, following from \eq{SORlightcone}. To order $1/\e^2$ in the $\overline{\text{MS}}$ scheme, the mixed-scale term is
\begin{align} \label{S2IO}
S_{\text{NG}}^{\text{OR}} &= \frac{\as(\mu)^2 C_F C_A}{(2\pi)^2} \frac{\left(\mu^2 e^{\gamma_E}/2 \right)^{2\e}}{\Gamma(1-\e)^2} \, 2f_{\text{OR}}(R_R, R_L) \nn \\
&\qquad \times \Lambda^{-1-2\e} k_R^{-1-2\e} \tan^{2\epsilon}\frac{R_R}{2} \,.
\end{align}
Similarly, for the NGL depending on both $k_L$ and $k_R$, the mixed scale term following from \eq{SLRlightcone} is
\begin{align} \label{S2LR}
S_{\text{NG}}^{\text{LR}} &= \frac{\as(\mu)^2 C_F C_A}{(2\pi)^2} \frac{\left(\mu^2 e^{\gamma_E} \right)^{2\e}}{\Gamma(1-\e)^2} \, 2f_{\text{LR}} (R_L, R_R) \nn \\
& \qquad \times ( k_L k_R)^{-1-2\e} \tan^{2\epsilon}\frac{R_L}{2} \tan^{2\epsilon}\frac{R_R}{2} \,.
\end{align}
The $1/\e$ poles in these contributions are infrared in origin.  Now, the full soft function is infrared finite.  As argued in \cite{Hornig:2011iu}, this means that the purely ``in'' and ``out'' contributions in \eq{doublelogexp} contribute compensating IR divergent terms that cancel the IR poles in \eqs{S2IO}{S2LR}. This also cancels the $\mu$-dependent terms in \eqs{S2IO}{S2LR}, preserving RG invariance of the factorized cross section. After this cancellation, double logs of $k_{L,R}/\Lambda$ and $k_L/k_R$ survive in the full soft function, and similar double logs of the other scale ratios survive.  These are the NGLs.

The constraints from RG invariance imply relationships between the NGLs in \eq{doubleNGLs}.  It is instructive to break up the contributions to the non-global double logs in terms of what regions the soft gluons are in.  There are six such regions, and all of the contributions have a coefficient $-(\alpha_s/2\pi)^2 C_F C_A$:
\begin{align} \label{NGLterms}
&f_{\text{L}} (R_L) \ln^2 \frac{\mu \tan(R_L/2)}{k_L} \,, \nn \\
&f_{\text{R}} (R_R) \ln^2 \frac{\mu \tan(R_R/2)}{k_R} \,, \nn \\
&f_{\text{O}} (R_L, R_R) \ln^2 \frac{\mu}{2\Lambda} \,, \\
-& f_{\text{OL}} (R_L, R_R) \ln^2 \frac{\mu^2 \tan(R_L/2)}{2\Lambda k_L} \,, \nn \\
-& f_{\text{OR}} (R_R, R_L) \ln^2 \frac{\mu^2 \tan(R_R/2)}{2\Lambda k_R} \,, \nn \\
-& f_{\text{LR}} (R_L, R_R) \ln^2 \frac{\mu^2 \tan(R_L/2) \tan (R_R/2)}{k_L k_R} \,. \nn
\end{align} 
Note that the first three coefficients receive contributions from purely real diagrams (with both soft gluons in the final state in the same region) and real-virtual diagrams (with one soft gluon in the final state and one virtual soft gluon).  There are several properties of these coefficients:
\begin{itemize} \label{NGLrelations}
\item $f_{\text{L}} = f_{\text{R}}$ ,
\item $f_{\text{O}} (R_L, R_R) = f_{\text{O}} (R_R, R_L)$ ,
\item $f_{\text{LR}} (R_L, R_R) = f_{\text{LR}} (R_R, R_L)$ ,
\item $f_{\text{OL}} (R_L, R_R) = f_{\text{OR}} (R_R, R_L)$ ,
\item $f_{\text{OL}}$ and $f_{\text{OR}}$ may not be symmetric in their arguments.
\end{itemize}
Finally, the statement that the NGLs are determined purely by RG invariance and the mixed scale logs ($f_{\text{LR}}$, $f_{\text{OL}}$, and $f_{\text{OR}}$) implies that there are relations between the coefficients.  Expanding the logs in \eq{doubleNGLs} and using \eq{doublelogexp},
\begin{subequations}
\label{relations}
\begin{align}
f_{\text{L}}(R_L) &= 2[f_{\text{LR}} (R_L, R_R) + f_{\text{OL}} (R_L, R_R)] \,,  \\
f_{\text{R}}(R_R) &= 2[f_{\text{LR}} (R_L, R_R) + f_{\text{OR}} (R_R, R_L)] \,,  \\
f_{\text{O}}(R_L, R_R) &= 2[f_{\text{OL}} (R_L, R_R) + f_{\text{OR}} (R_R, R_L)] \,.
\end{align}
\end{subequations}
So far we have not used that $R_{L,R}$ could be different, so our proof of \eq{relations} is valid for equal $R$'s. Now, we can use different $R_{L,R}$ to argue that the $R_L$ dependence cancels between $f_{\text{LR}}$ and $f_{\text{OR}}$; similarly, the $R_R$ dependence cancels between $f_{\text{LR}}$ and $f_{\text{OL}}$.  If $f_{\text{OR}}$ is known, then up to a constant $f_{\text{LR}}$ can be determined.  We have the additional constraint that as the jet radius shrinks to zero, the NGL of the two jet scales will also vanish: $f_{\text{LR}} (R_L, R_R) \to 0$ as $R_L \to 0$ or $R_R \to 0$.  This means that only knowing $f_{\text{OR}}$ completely determines all the other coefficients.  We will compute $f_{\text{OR}}$ for the anti-\kt algorithm and use it to determine $f_{\text{LR}}$.  For the C/A and \kt algorithms we will take $R_L = R_R = R$.

\section{Non-Global Logs for Several Algorithms}
\label{sec:NGL}

In this section we derive results for the non-global part $S_{\text{NG}}$ of the soft function in \eq{factorizedsoft} not predicted by RG.  For each algorithm we first determine the ``in-out'' NGLs by calculating $f_{\text{OR}}$, and then determine the ``in-in'' NGLs by calculating $f_{\text{LR}}$.  After calculating the leading NGLs for each algorithm, we plot the coefficients of the logs and discuss the results.

As shown in \sec{sec:doubleNGLs}, the double log terms in $S_{\text{NG}}$ can be determined by the calculation of $f_{\text{OR}}$, the contribution with one gluon in a jet (in this case the right jet) and one gluon out of the jets. As is well known, at $\ca{O}(\as^2)$ the non-global double logs arise from soft gluon emission diagrams with the $C_F C_A$ color structure, with the amplitude in \eq{NGLamplitude}.

From the form of \eqs{SORlightcone}{SLRlightcone} one finds that the coefficients of the leading NGLs are given generically by the integral (cf. \cite{Appleby:2002ke})
\begin{align}
\label{fOR}
f_{\text{OL,OR,LR}} = 2\int_{-\infty}^\infty \!\! d\eta_1 d\eta_2 \!\! \int_0^\pi \! \frac{d\phi}{\pi} & \frac{\cos\phi}{\cosh(\eta_1 - \eta_2)  - \cos\phi} \nn \\
&\times\Theta_{\text{OL,OR,LR}}^{\rm alg} \,,
\end{align}
where $\eta_{1,2} = \ln\cot(\theta_{1,2}/2)$ are the (pseudo-)rapidities of gluons $1,2$ with respect to the $z$ axis (the jet 1 axis).  The angular constraints of the jet algorithm are given in $\Theta_{\rm alg}$, and depend on which coefficient (OL, OR, or LR) we are calculating.  
This integral is a geometric measure of the size of the region into which the two soft gluons are allowed to go for a given contribution to the NGL.

We consider two types of jet algorithms, cone and recombination.  Cone algorithms fit jets to a geometric shape (the cone), and a jet is found when the momentum in the cone is aligned with its axis.  Therefore, the phase space constraints for particles in a found jet simply requires them to be within an angle $R$ of the cone axis.  For soft particles in the $n$ jet, for instance, this implies $\theta_{ns} < R$.

Recombination algorithms build jets by successive $2\to1$ mergings of particles.  A pairwise metric $d_{ij}$ and a single particle metric $d_i$ govern the recombinations.  A single step in the algorithm finds the smallest of all $d_{ij}$ and $d_i$, then merges the closest pair if some $d_{ij}$ is smallest or promotes a particle to a jet if some $d_i$ is the smallest.  This procedure is repeated until all particles are put into jets.  The common recombination algorithms (\kt, C/A, and anti-\kt) are part of a class parameterized by a real number $\alpha$.  In terms of $\alpha$, the metrics for $e^+e^-$ are
\begin{align}
d_{ij} &= 2\min(E_i^{2\alpha}, E_j^{2\alpha}) (1 - \cos\theta_{ij}) \,, \nn \\
d_i &= 2E_i^{2\alpha} (1-\cos R) \,.
\end{align}
For both types of algorithms, a veto $\Lambda$ on soft jets is required for infrared safety of exclusive jet cross sections.

\subsection{Cone or anti-\kt~algorithms}

To leading order in the SCET power counting, the phase space for soft particles that get combined into the jets is the same in the cone and anti-\kt~algorithms \cite{Ellis:2009wj,Ellis:2010rw}, so they will have the same leading NGL (also pointed out in \cite{Cacciari:2008gp}). We work with fixed cones, but the results apply to other infrared-safe cone algorithms (e.g. \cite{Salam:2007xv}) in the configurations we consider where there is no split-merge issue. For generic jet configurations, different cone algorithms will  have different NGLs.

\subsubsection{In-Out NGLs}

For these algorithms, the two-particle phase space for one soft parton to be inside and one outside of a jet is
\begin{align} \label{PScone}
\Theta_{\text{OR}}^{\text{cone}} &= \theta(\eta_R<\eta_1< \infty)\, \theta(-\eta_L < \eta_2 < \eta_R) \nn \\ 
& \qquad + (1\leftrightarrow2) \, ,
\end{align}
where $\eta_{L,R} = \ln \cot R_{L,R}/2$.  Interchanging the gluons simply introduces a factor of 2 into the integral, and so we will simply work with the first configuration shown in \eq{PScone} and multiply by 2 to account for this symmetry.  
With these constraints the in-out NGL coefficient $f_{\text{OR}}(R_R,R_L)$ in \eq{fOR} is then given by
\be
\label{fR}
f_{\text{OR}}^{\text{cone}}(R_R,R_L) = \int_{\eta_R}^\infty d\eta_1 \int_{-\eta_L}^{\eta_R}d\eta_2 \frac{8}{e^{2(\eta_1-\eta_2)} - 1} \,,
\ee
The integrand depends only on the difference $\eta_1-\eta_2$, and can be easily integrated to give
\be
\label{fRcone}
\begin{split}
&f_{\text{OR}}^{\text{cone}}(R_R,R_L) = \frac{\pi^2}{3}  - 2\Li_2\left(\! \tan^2\frac{R_L}{2} \tan^2\frac{R_R}{2}\!\right) \,.
\end{split}
\ee
In \fig{fig:fOR} we plot \eq{fRcone} for $R_L=R_R\equiv R$. At $R=0$, $f_{\text{OR}}^{\text{cone}}(R)\to \pi^2/3$, and at $R\to \pi/2$ (hemisphere jets), $f_{\text{OR}}^{\text{cone}}(R)\to 0$.

This result is consistent with that of \cite{Banfi:2010pa} in the limit $R\to 0$, who considered the case of measuring only one jet's invariant mass and imposing a jet veto outside the two jets. Adding two copies of the NGL they found in that case  reproduces \eq{fRcone} (for $R\to 0$). \eq{fRcone} now provides the full $R$ dependence of the coefficient of the in-out NGLs in \eqs{OLNGL}{ORNGL} for the cone or anti-\kt algorithms, with the separate $R_{L,R}$ dependence derived for the first time.\footnote{Ref.~\cite{Banfi:2002hw} calculated the same coefficient \eq{fRcone} for the NGL of the jet veto over the total energy in two-jet events in which the jet masses are not measured (for $R_L=R_R=R$).}

\subsubsection{In-In NGL}

We can use the constraints in \sec{sec:doubleNGLs} to determine the in-in NGL coefficient, $f_{\text{LR}}^{\text{cone}}$.  The sum $f_{\text{OR}} + f_{\text{LR}}$ must be $R_L$ independent, and $f_{\text{LR}}$ is symmetric in its arguments and vanishing as $R_{L,R}\to0$.  Since $f_{\text{OR}}^{\text{cone}}$ in \eq{fRcone} happens to be symmetric in $R_{L,R}$, it is simple to determine
\be \label{fLRcone}
f_{\text{LR}} (R_L, R_R) = 2\Li_2\left(\! \tan^2\frac{R_L}{2} \tan^2\frac{R_R}{2}\!\right) \,.
\ee
This result is simple to check by performing the integral in \eq{fOR} with the constraint
\be
\Theta_{\text{LR}}^{\text{cone}} = \theta(\eta_R < \eta_1 < \infty) \, \theta(-\infty < \eta_2 < \eta_L) \,,
\ee
which confirms the result in \eq{fLRcone}:
\begin{align} \label{fLRconeintegral}
f_{\text{LR}}^{\text{cone}}(R_L,R_R) &=  \int_{\eta_R}^\infty  d\eta_1 \int_{-\infty}^{-\eta_L} d\eta_2 \frac{8}{e^{2(\eta_1-\eta_2)}-1} \,, \nn \\
&= 2\Li_2\left(\! \tan^2\frac{R_L}{2} \tan^2\frac{R_R}{2}\!\right) \,.
\end{align}
In \fig{fig:fLR} we plot this coefficient for $R_L = R_R = R$.  

The sum of the in-out and in-in coefficients is related to the coefficient $f_R$, which comes from having gluons only in the right jet:
\be
f_{\text{OR}}^{\text{cone}} (R_R,R_L) + f_{\text{LR}}^{\text{cone}} (R_L,R_R) = \frac{1}{2}f_{\text{R}}^{\text{cone}} (R_R) = \frac{\pi^2}{3} \,.
\ee
The fact that $f_{\text{R}}^{\text{cone}}$ is a constant independent of $R_R$ is somewhat surprising, but can be shown at the level of the integrals in \eqs{fR}{fLRconeintegral}.  These integrals sum simply to give
\be
f_{\text{R}}^{\text{cone}} (R_R) = \int_{\eta_R}^{\infty} d\eta_1 \int_{-\infty}^{\eta_R} d\eta_2 \frac{16}{e^{2(\eta_1 - \eta_2)} - 1} \,.
\ee
The integrand depends only on $\eta_1 - \eta_2$, and we can shift the integration variables by $\eta_{1,2} \to \eta_{1,2} - \eta_R$, rendering the integral to be a constant:
\be \label{fRakt}
f_{\text{R}}^{\text{cone}} (R_R) = \int_{0}^{\infty} d\eta_1 \int_{-\infty}^{0} d\eta_2 \frac{16}{e^{2(\eta_1 - \eta_2)} - 1} = \frac{2\pi^2}{3} \,.
\ee
In \fig{fig:fR}, we compare this constant to the result for $f_R^{\text{alg}}$ from other algorithms.  Ref.~\cite{Rubin:2010fc} also noted that $f_{\text{R}}^{\text{cone}}$ is a constant but noted that no physical explanation was apparent. We observe above that the boost invariance of the amplitude and the property that all rapidities get covered in the sum $f_{\text{LR}}^{\text{cone}}+f_{\text{OR}}^{\text{cone}}$ removes dependence on where precisely that jet boundary $\eta_R$ is. Below we will find that this coefficient is not a constant for the C/A and \kt algorithms, where the phase space for recombination of two soft gluons in a single jet is distinct from anti-\kt and changes the value of $f_{\text{R}}$. In particular, the phase space included in the sum $f_{\text{LR}}+f_{\text{OR}}$ contains gaps not invariant under boosts so that dependence on the boundaries remains.

\subsection{Cambridge-Aachen algorithm}

In computing the leading NGLs for the C/A and \kt algorithms, we will use only one common jet radius, setting $R_L = R_R = R$.  In these cases $f_{\text{OR}} = f_{\text{OL}}$ and $f_{\text{R}} = f_{\text{L}}$.

\subsubsection{In-Out NGLs}

The phase space constraints from the C/A algorithm that figure into $f_{\text{OR}}$ are more complicated than the cone or anti-\kt algorithms.  To contribute to $f_{\text{OR}}$, one of the soft gluons must be merged into the right jet while the other is not recombined with either jet.  This amounts to the phase space constraints
\begin{align}
\Theta_{\text{OR}}^{\text{C/A}} &= \theta(0< \theta_1 < R)\, \theta(R<\theta_2<\pi-R) \nn \\
& \qquad \times \theta(\theta_1 < \theta_{12}) \,.
\end{align}
Here $\theta_{12}$ is the opening angle between gluons 1 and 2, while $\theta_1$ and $\theta_2$ are the angles between each gluon and the right jet axis.  The last condition is required to guarantee that partons 1 and 2 are not combined together first by the algorithm, in which case they would be in the ``in-in'' or ``out-out'' part of the phase space not contributing to $f_{\text{OR}}^{\text{C/A}}$.  An analogous constraint only contributes at the level of a power correction for the cone or anti-\kt algorithms.

The coefficient $f_{\text{OR}}^{\text{C/A}}$ is then determined by the integral
\begin{align} \label{fCA}
f_{\text{OR}}^{\text{C/A}}(R) &= 4\int_{\eta_R}^\infty  d\eta_1 \int_{-\eta_R}^{\eta_R} d\eta_2\int_0^\pi \frac{d\phi}{\pi} \ \Theta_{\text{OR}}^{\text{C/A}} \nn \\
&\quad\times\frac{\cos\phi}{\cosh(\eta_1-\eta_2) - \cos\phi} \,.
\end{align}
Here $\eta_R = \ln \cot R/2$.  The phase space constraints in $\Theta_{\text{OR}}^{\text{C/A}}$ can be written as 1 minus the region where the soft partons $1,2$ \emph{do} get combined by the C/A algorithm,
\begin{align} \label{ThetaCA}
\theta(\theta_{12} & -\theta_1) = 1 - \Theta(\theta_1 - \theta_{12}  ) \\
&=  1 -  \Theta\left(\cos\phi > \max\{e^{-\eta_2}\sinh \eta_1, -e^{\eta_1}\sinh\eta_2\}\right) \nn \,.
\end{align}
The last theta function enforces that partons 1 and 2 are closer to each other in angle than the jet axes. The ``1'' term in \eq{ThetaCA} produces the same result as $f_{\text{OR}}^{\text{cone}}$, so that \eq{fCA} can be written as
\be
\label{fROCA}
f_{\text{OR}}^{\text{C/A}}(R) = f_{\text{OR}}^{\text{cone}}(R) - \Delta f_{\text{OR}}^{\text{C/A}}(R)\,.
\ee
To evaluate $\Delta f_{\text{OR}}^{\text{C/A}}$, we can perform the integral over $\phi$ using (for $\eta>0$)
\begin{align}
\int d\phi & \,\frac{\cos\phi}{\cosh\eta - \cos\phi} \nn \\
& \qquad = -\phi + \frac{2}{\tanh\eta}\tan^{-1}\!\left[\frac{\tan(\phi/2)}{\tanh(\eta/2)}\right] \,,
\end{align}
and evaluate the integrals over $\eta_{1,2}$ numerically.  The result is plotted in \fig{fig:fOR}.

\subsubsection{In-In NGLs}

The contribution to $f_{\text{LR}}^{\text{C/A}}$ requires one gluon in each jet.  The two gluons must merge with the jets before merging with each other, which requires $\theta_{12} > \max(\theta_1,\theta_2)$.  Therefore
\begin{align}
\Theta_{\text{LR}}^{\text{C/A}} &= \theta(0< \theta_1 < R)\, \theta(\pi-R<\theta_2<\pi) \nn \\
& \qquad \times \theta( \max(\theta_1,\theta_2) < \theta_{12} ) \,.
\end{align}
Again we can divide up these constraints into a contribution identical to the cone restrictions and a correction factor,
\be
\label{fLRCA}
f_{\text{LR}}^{\text{C/A}}(R) = f_{\text{LR}}^{\text{cone}}(R) - \Delta f_{\text{LR}}^{\text{C/A}}(R) \,,
\ee
where the last term is given by
\begin{align} \label{fLRCAintegrals}
& \Delta f_{\text{LR}}^{\text{C/A}}(R) = \frac{8}{\pi}\int_{\eta_R}^{\eta_{R/2}} d\eta_1 \!  \int_{-\eta_1}^{-\eta_R}  \!\! d\eta_2 \nn \\
 &\times  \Theta\left(\eta_2 - \ln\sinh\eta_1\right) \Biggl\{ - \cos^{-1}\left(e^{-\eta_2}\sinh\eta_1\right) \\
 &\plus \frac{2}{\tanh(\eta_1\minus \eta_2)} \! \tan^{-1}\!\!\left[\coth\left(\!\frac{\eta_1\minus\eta_2}{2}\!\right)\! \!\sqrt{\frac{1\minus e^{-\eta_2}\sinh\eta_1}{1\plus e^{-\eta_2}\sinh\eta_1}}\right] \!\Biggr\}. \nn
\end{align}

Considering the phase space constraints, it is straightforward to see that if $R < \pi/3$, the in-in NGL coefficients must be the same for cone, anti-\kt, and C/A (as well as \kt).  This is because for a sufficiently small jet radius, the soft gluons cannot merge together since $\theta_{12} > R$.  This is confirmed by the calculation of $f_{\text{LR}}^{\text{C/A}}$, and is shown in \fig{fig:fLR}.

With the results for $f_{\text{OR}}^{\text{C/A}}$ and $f_{\text{LR}}^{\text{C/A}}$, we can combine them to determine $f_{\text{R}}^{\text{C/A}}$.  Unlike the anti-\kt algorithm, this coefficient is $R$ dependent for the C/A algorithm, and is plotted in \fig{fig:fR}. 

\subsection{\kt~algorithm}

The phase space constraints from \kt algorithm are more complex than the C/A algorithm, as there are different limiting cases to consider where the NGLs can be large.  To determine whether two soft gluons are recombined by the \kt algorithm or not, we consider the pairwise distances among the soft particles and the two collinear jets,
\begin{align}
d_{12} &= 2\min(E_1^2,E_2^2) \, (1 - \cos\theta_{12}) \,, \nn \\
d_{in} &= 2E_i^2 \, (1 - \cos\theta_i)\, , \quad d_{i\bn} = 2E_i^2 \, (1 + \cos\theta_i) \,,
\end{align}
for $i = 1,2$.  The single particle metrics are
\be
d_i = 2E_i^2 (1 - \cos R) \,.
\ee
The phase space constraints from the \kt algorithm depend on the relative scaling of the two soft gluons.  For instance, if gluon 1 is in the $R$ jet and gluon 2 is out of both jets, then the limit $E_1 \ll E_2$ will give rise to a different phase space than the opposite limit, $E_2 \ll E_1$.  Both limits will give rise to a large NGL, and we will consider both regimes in calculating the coefficients for the different logs.  The limit $E_1 \ll E_2$ will produce the same phase space constraints as the C/A algorithm, meaning that the NGLs will be the same for \kt and C/A in this limit.  In the limit $E_2 \ll E_1$, we find a new coefficient for the NGLs.

\subsubsection{In-Out NGLs}

As with the previous algorithms, the coefficient $f_{\text{OR}}^{\text{\kt}}$ is given by
\begin{align} \label{fkT}
f_{\text{OR}}^{\text{\kt}}(R) &= 4\int_{\eta_R}^\infty  d\eta_1 \int_{-\eta_R}^{\eta_R} d\eta_2\int_0^\pi \frac{d\phi}{\pi} \ \Theta_{\text{OR}}^{\text{\kt}} \nn \\
&\quad\times\frac{\cos\phi}{\cosh(\eta_1-\eta_2) - \cos\phi} \,.
\end{align}
The phase space restrictions are given in general by
\begin{align} \label{kTconstraints}
\Theta_{\text{OR}}^{\text{\kt}} &= \theta(0 < \theta_1 < R) \, \theta(R < \theta_2 < \pi - R) \nn \\
& \quad \times \theta( \min(d_{1n},d_2) < d_{12}) \,.
\end{align}
The first two theta functions require that $d_{1n} < d_1$ and $d_2 < \{ d_{2\bn}, d_{2n}\}$, while the last requires that the two soft gluons are not merged together by the algorithm.

The ordering in the last theta function in \eq{kTconstraints} depends on the soft gluon kinematics.  We consider two limits, both of which give a large NGL coefficient: $k_R \ll \Lambda \tan R/2$ and $\Lambda \tan R/2 \ll k_R$.  

If $k_R \ll \Lambda \tan R/2$, then
\be
E_1 (1-\cos\theta_1) \ll E_2 \tan R/2 \,.
\ee
Unless $\theta_1 \ll 1$ (which is a power correction for the contribution to the NGL coefficient), then this limit implies $d_{1n} < d_2$.  This means the relevant comparison is $d_{1n} < d_{12}$, which simplifies to the same constraint as the C/A algorithm,
\be \label{kTisCA}
\theta_1 < \theta_{12} \,.
\ee
In this region the coefficient $f_{\text{OR}}$ is the same as for the Cambridge/Aachen algorithm.
\begin{figure}[t]{
\centering
\includegraphics[width=.99\columnwidth]{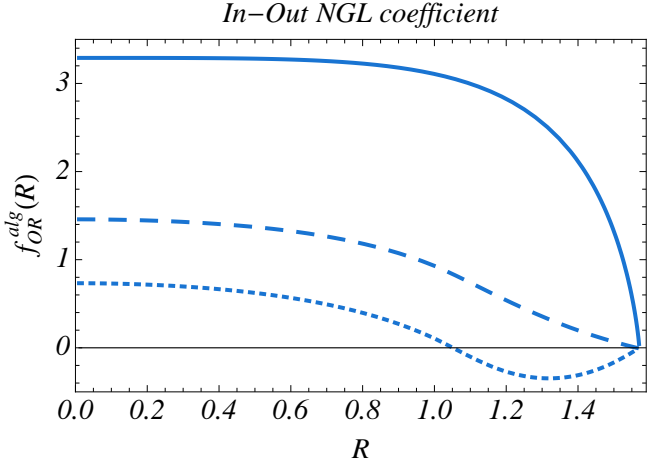}
\vspace{-1em}
{ \caption[1]{``In-Out'' NGLs for three algorithms. The coefficient $f_{\text{OR}}^{\text{alg}}(R)$ (equivalently $f_{\text{OL}}^{\text{alg}}$) of the leading NGL $\ln^2[k_R/(\Lambda\tan R/2)]$ for the cone/anti-\kt algorithms (solid), the Cambridge-Aachen algorithm (dashed), and the \kt\ algorithm when $k_R\ll \Lambda$ (also dashed) and when $\Lambda\ll k_R$ (dotted). These algorithms  recombine soft gluons in successively larger regions of phase space, reducing the coefficient of the NGL.}
\label{fig:fOR}}
}
\end{figure}

In the opposite limit of the NGL scales, $\Lambda \tan R/2 \ll k_R$,
\be
E_2 \tan R/2 \ll E_1 (1-\cos\theta_1) \,.
\ee
Outside of the power suppressed region of phase space, $d_2 < d_{1n}$, and the constraint becomes $d_2 < d_{12}$.  This is more restrictive than the C/A case, equivalent to
\be \label{kTphasespace}
R < \theta_{12} \,.
\ee
In this region the coefficient of the NGL of $\Lambda/k_R$ can be written similarly to \eq{fROCA}, 
\be
\label{fROkT}
f_{\text{OR}}^{\text{\kt}}(R) = f_{\text{OR}}^{\text{cone}}(R) - \Delta f_{\text{OR}}^{\text{\kt}}(R)\,,
\ee
where $\Delta f_{\text{OR}}^{\text{\kt}}$ is given by
\begin{align} \label{DeltafROkT}
\Delta f_{\text{OR}}^{\text{\kt}}(R) &= 4\int_{\eta_R}^\infty d\eta_1 \int_{-\eta_R}^{\eta_R}d\eta_2 \int_0^\pi \frac{d\phi}{\pi} \nn \\
&\quad\times  \frac{\cos\phi}{\cosh(\eta_1\minus\eta_2) - \cos\phi}  \theta(R - \theta_{12})\,,
\end{align}
whose $\phi$ integral can be evaluated to give
\begin{align} \label{DeltafROkTint}
\Delta f_{\text{OR}}^{\text{\kt}}(R)  &= 4\int_{\eta_R}^\infty d\eta_1 \int_{\log\cot[\max(\pi-R,\theta_1+R)]}^{\eta_R} d\eta_2  \\ 
&\times \left\{\! -\phi_0 + \frac{2}{\tanh(\eta_1\minus \eta_2)} \tan^{-1}\!\!\left[\frac{\tan(\phi_0/2)}{\tanh\frac{\eta_1-\eta_2}{2}}\right]\! \right\} , \nn
\end{align}
where
\be
\phi_0 = \cos^{-1}\! \left[\cos^2\frac{R}{2} \cosh(\eta_1\minus \eta_2) - \sin^2\frac{R}{2}\cosh(\eta_1\plus\eta_2)\right] ,
\ee
and $\theta_1 = 2\cot^{-1}e^{\eta_1}$.
We perform the remaining two integrals in \eq{DeltafROkTint} numerically and plot the result in \fig{fig:fOR}.

\subsubsection{In-In NGLs}

\begin{figure}[t!]{
\centering
\includegraphics[width=\columnwidth]{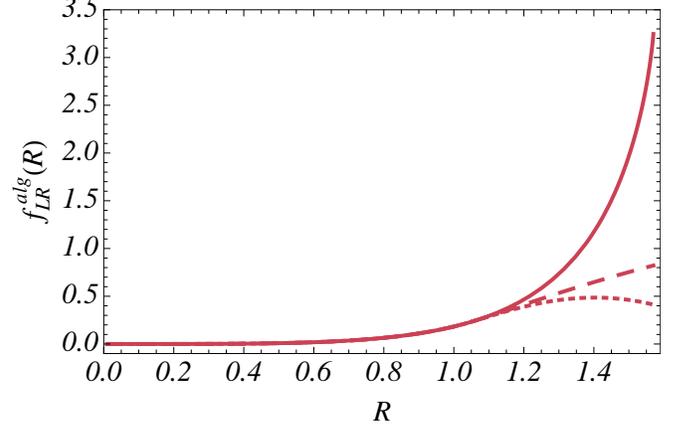}
\vspace{-1em}
{ \caption[1]{``In-In'' NGLs for three algorithms. The coefficient $f_{\text{LR}}^{\text{alg}}(R)$ of the leading NGL $\ln^2(k_L/k_R)$ for the cone/anti-\kt algorithms (solid), the Cambridge-Aachen algorithm (dashed), and the \kt\ algorithm in the limit $k_L\ll k_R$ (or $k_R\ll k_L$) (dotted). The coefficients only differ for $R>\pi/3$, the smallest angle for which the algorithms can begin to recombine soft gluons in separate jets.}
\label{fig:fLR}}
}
\end{figure}

\begin{figure}[t!]{
\centering
\includegraphics[width=\columnwidth]{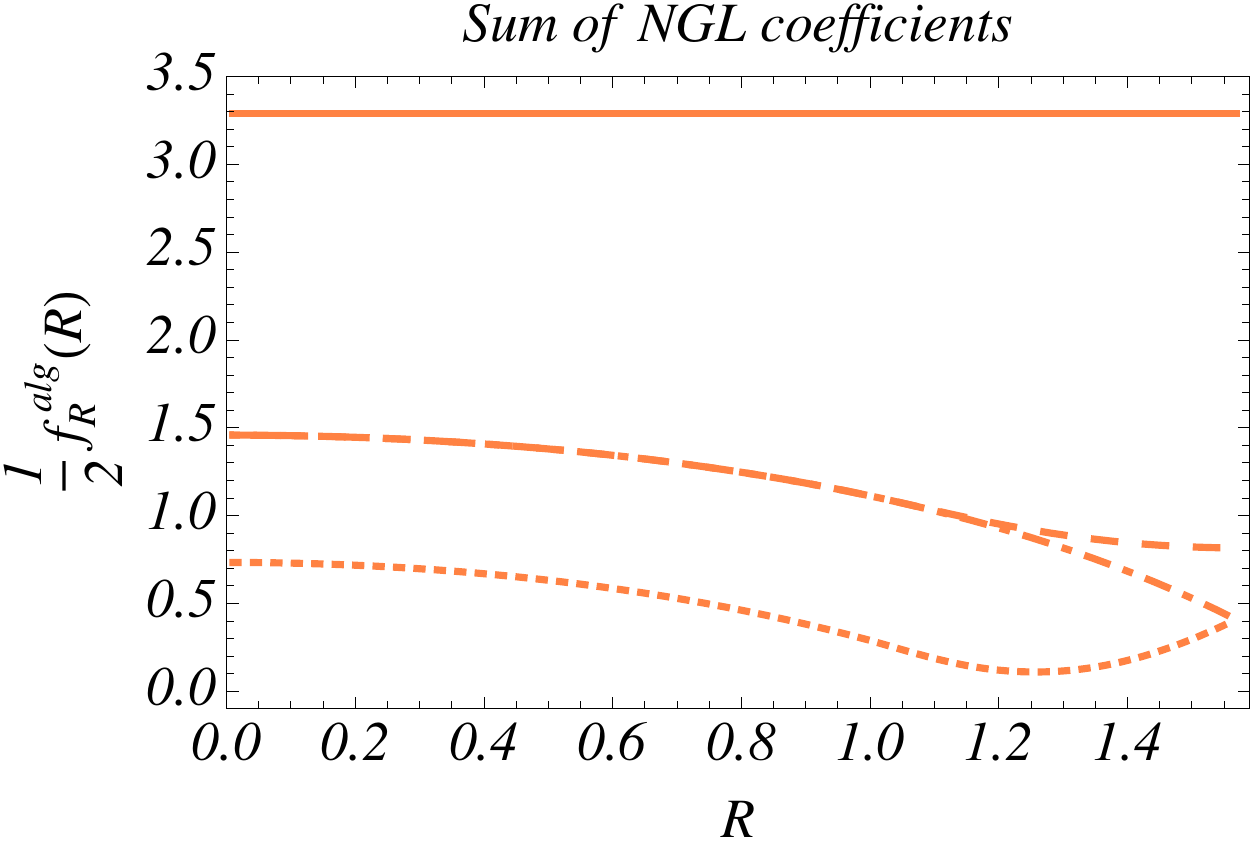}
\vspace{-1em}
{ \caption[1]{Sum of NGL coefficients for three algorithms. As derived in \sec{sec:doubleNGLs}, the sum $f_{\text{OR}}+f_{\text{LR}}$ is equal to $\frac12 f_R$, the contribution to NGLs from gluons just in the R jet. We plot $\frac12 f_R^{\text{alg}}$ for the cone/anti-\kt (solid),  Cambridge-Aachen (dashed), and  \kt\ algorithms when $k_R\ll \Lambda$ (dot-dashed) and when $\Lambda\ll k_R$ (dotted). $\frac12 f_R^{\text{cone}}$ is a constant, $\pi^2/3$.}
\label{fig:fR}}
}
\end{figure}

When one soft gluon is inside each of the two jets, we require
\be
d_{1n} \,, d_{2\bn} < d_{12} \,.
\ee
As with $f_{\text{OR}}^{\text{\kt}}$, we want to consider the limits of a large NGL, namely $k_L \ll k_R$ and $k_R \ll k_L$.  We can consider these limits simultaneously, taking $k_R \ll k_L$ for definiteness.  At leading power this implies $E_1 \ll E_2$ and $d_{1n} \ll d_{2\bn}$, since we can neglect the small angle region near the jet axes.  The constraint therefore becomes
\be
\theta_{1n} < \theta_{12} \,,
\ee
and we have
\begin{align}
\Theta_{\text{LR}}^{\text{\kt}} = \theta(0 \!<\! \theta_1 \!<\! R)\, \theta(\pi\minus R\!<\!\theta_2<\pi) \theta( \theta_1 \!< \!\theta_{12} ) \,.
\end{align}
When $k_L \ll k_R$, the last constraint changes to $\theta_2 < \theta_{12}$, but this results in an identical coefficient.  As before, we can divide the calculation as
\be
\label{fLRkT}
f_{\text{LR}}^{\text{\kt}}(R) = f_{\text{LR}}^{\text{cone}}(R) - \Delta f_{\text{LR}}^{\text{\kt}}(R)\,,
\ee
where
\begin{align} \label{DeltafLRkT}
\Delta f_{\text{LR}}^{\text{\kt}}(R) &= 4\int_{\eta_R}^\infty \!\! d\eta_1 \int_{-\infty}^{-\eta_R}\!\! d\eta_2 \int_0^\pi \! \frac{d\phi}{\pi} \frac{\cos\phi}{\cosh(\eta_1\minus\eta_2) \minus \cos\phi} \nn \\
&\qquad\times \theta(\theta_1 - \theta_{12})\,,
\end{align}
where 
\begin{align}
&\Delta f_{\text{LR}}^{\text{\kt}}(R)  = 4\int_{\eta_R}^\infty d\eta_1 \int_{\min(-\eta_R,\ln\sinh\eta_1)}^{-\eta_R} d\eta_2 \nn \\ 
&\times \Biggl\{ - \cos^{-1}\left(e^{-\eta_2}\sinh\eta_1\right) \\
 &+\! \frac{2}{\tanh(\eta_1\minus \eta_2)} \tan^{-1}\!\!\left[\coth\!\left(\!\frac{\eta_1\minus\eta_2}{2}\!\right)\!\! \sqrt{\frac{1\minus e^{-\eta_2}\sinh\eta_1}{1\plus e^{-\eta_2}\sinh\eta_1}}\right]\! \Biggr\} . \nn
\end{align}
We perform the remaining integrals numerically and plot the result in \fig{fig:fLR}.  We also sum the in-out and in-in coefficients, $f_{\text{OR}}^{\text{\kt}} + f_{\text{LR}}^{\text{\kt}} = \frac{1}{2}f_{\text{R}}^{\text{\kt}}$, and plot it in \fig{fig:fR}.

\subsection{Leading NGLs}

With the leading NGLs in hand, we can compare the coefficients between algorithms and learn about the structure of the NGLs.  These observations confirm and extend previous studies of clustering effects on NGLs (e.g. \cite{Appleby:2002ke,Delenda:2006nf,Banfi:2010pa}). Our extensive investigation of the algorithm and $R$ dependence found above makes clearer this connection between clustering algorithms and NGLs and allows us to make more specific statements about the field theoretic origin and properties of NGLs.

We first build a picture of the NGLs that we can use to interpret the results for different algorithms.  For all algorithms, the matrix element contributing to the coefficients $f_{\text{OR}}$, $f_{\text{OL}}$, and $f_{\text{LR}}$ is proportional to
\be \label{collinearenhancement}
\frac{\cos\Delta\phi}{\cosh(\Delta \eta) - \cos\Delta\phi} \,,
\ee
where $\Delta\phi$ is the azimuthal angle difference and $\Delta\eta$ is the pseudorapidity difference between the soft gluons.  The phase space restrictions require the gluons to be in different regions, so that $\Delta\eta > 0$.  The region $\Delta\eta$ and $\Delta\phi$ near zero provides a collinear enhancement to the NGL coefficients, but there is no collinear singularity that sets the value of the coefficient.  This implies that the NGLs are soft logarithms.  The infrared singularities that are contained in the distributions
\be
k_{L,R}^{-1-2\e} \quad \textrm{ and } \quad \Lambda^{-1-2\e} 
\ee
come from the soft region of phase space, when $E_{1,2} \to 0$.  The coefficients of the NGLs receive support over the entire region of phase space, with the dominant contribution near the jet boundary.  Since different jet algorithms merge nearby soft gluons in different ways, the coefficient of the in-out NGLs is very different.  We now discuss the results for each algorithm.

We start with the in-out coefficient, $f_{\text{OR}}^{\text{alg}} (R)$ (equivalently $f_{\text{OL}}^{\text{alg}}$).  In \fig{fig:fOR}, we plot this coefficient as a function of $R$.  The cone and anti-\kt coefficient is significantly larger than either C/A or \kt.  This is a manifestation of the in-out phase space.  For the cone and anti-\kt algorithms, the soft gluons on either side of the jet boundary will not recombine together.  The soft gluons that are close to the jet boundary (one in the jet, one out) contribute more to the NGL from a collinear enhancement in the matrix element (see \eq{collinearenhancement}).  For the C/A and \kt algorithms, soft gluons near the boundary will recombine together and not contribute to $f_{\text{OR}}$.  This removes the region with collinear enhancement and subsequently reduces the size of the coefficient.  Since the \kt algorithm also weights the pairwise recombination metric by the minimum energy of the pair, the region of phase space for soft gluon recombination is larger.  This further reduces the size of the coefficient, and for large $R$ it even changes sign.  For each algorithm, it is interesting to see that the NGL coefficient is nearly constant over a wide range of $R$; for the anti-\kt algorithm, the corrections for small $R$ go as $R^4$.

The behavior of the in-in coefficient is very different from the in-out coefficient.  The in-in coefficient, $f_{\text{LR}}^{\text{alg}} (R)$, for each algorithm is plotted in \fig{fig:fLR}.  Each algorithm gives the same coefficient for $R < \pi/3$, and the coefficients differ for larger $R$ values.  This comes from the action of the algorithm: for $R < \pi/3$, the two gluons are separated by more than $R$, and so they cannot be recombined.  Furthermore, since each jet contains only one soft gluon, the action of each algorithm is the same, and so $f_{\text{LR}} (R)$ is the same for each algorithm.

Note that as $R\to 0$, the phase space shrinks to 0 and the coefficient vanishes.  This supports the picture that the logs proportional to $f_{\text{LR}}$ are soft logarithms, since we would expect collinear logarithms to have a non-vanishing coefficient in the small $R$ limit.  The small size of $f_{\text{LR}}$ away from the small $R$ limit comes from the fact that the soft gluons must be separated by an angle greater than $\pi - 2R$.  This cuts out the region of phase space with a collinear enhancement in the coefficient and reduces its size.  As $R$ increases, the soft gluons can be recombined across the jets, which also reduces the magnitude of the NGL coefficient.

Finally, in \fig{fig:fR} we plot the sum of the NGL coefficients, $\frac{1}{2} f_{\text{R}}^{\text{alg}} (R) = f_{\text{OR}}^{\text{alg}} (R) + f_{\text{LR}}^{\text{alg}} (R)$ (or equivalently, $\frac{1}{2}f_{\text{L}}^{\text{alg}}$).  For the \kt algorithm, we have plotted the coefficient in the two limits $\Lambda \ll  Q\rho$ and $Q\rho \ll\Lambda$.  For the anti-\kt algorithm, this coefficient is a constant for all $R$.  This comes from the boost invariance properties of the matrix element that make it independent of the jet radius, see \eq{fRakt}.  Since $f_{\text{LR}}$ is the same for all algorithms for $R < \pi/3$ and the $R$ dependence of $f_{\text{OR}}$ is different for each algorithm, this implies that $f_{\text{R}}$ cannot be a constant for C/A or \kt.  The coefficient $f_{\text{R}}$ is the contribution to the NGLs from the region of phase space with soft gluons only in the (right) jet.  This coefficient can receive contributions from real configurations with two soft gluons in the final state or real-virtual configurations with one soft gluon in the final state and one virtual gluon.  Therefore the fact that the anti-\kt coefficient is constant is  possibly due to an accidental cancellation.  The fact that $f_{\text{R}}$ coefficient does not vanish as $R\to0$ indicates that at least part of the coefficient receives contributions from a collinear log.

\section{Comparison of Global and Non-Global Logs to EVENT2}
\label{sec:comparison}

We can test for the presence of the NGL of $\Lambda/m_{1,2}$ in $\sigma(m_1,m_2,\Lambda)$ implied by \eq{doubleNGLs} by comparing predictions with and without it to the output of EVENT2 \cite{Catani:1996jh,Catani:1996vz}.
To focus on this NGL let us consider measuring the two jets' \emph{total} invariant mass $\rho \equiv (m_1^2 + m_2^2)/ Q^2$. This prevents NGLs of $m_1/m_2$ from contributing.  But first, we must construct the prediction for the global logs of $\rho$ and $\Lambda/Q$ in the cross section.

\subsection{Global Logs}

The resummed cumulant cross section,

\begin{align} \label{cumulant}
\Sigma(\rho,\Lambda) = &\int_{-\infty}^{\rho}\!\! d\rho' \int \! dm_1^2 dm_2^2 \, \delta\left(\rho' - \frac{m_1^2 + m_2^2}{Q}\right)  \nn \\
&\times \int_{-\infty}^\Lambda d\Lambda'   \sigma(m_1,m_2,\Lambda'),
\end{align}
also splits into pieces predicted by RG evolution and those that are not and contain the NGLs,
\be
\label{cumulantfactorization}
\Sigma(\rho,\Lambda) = \Sigma_{\text{in}} (\rho)  \Sigma_{\text{out}} (\Lambda) \Sigma_{\text{NG}}(\rho,\Lambda)\,.
\ee
Predictions for the ``global'' pieces predicted by RG evolution, $\Sigma_{\text{in}}$ and $\Sigma_{\text{out}}$, are derived in \appx{appx:global}. 
\begin{widetext}
For the in-jet contribution,
\be
\label{rhoglobal}
\begin{split}
&\Sigma_{\text{in}}(\rho) = 1 \minus \frac{\as C_F}{\pi}\biggl[L_\rho^2 + \left(\frac{3}{2} \minus 4L_R\right)\! L_\rho  + \frac{1}{2} \minus \frac{\pi^2}{6} + 2L_R^2 \biggr] \\
& + \left(\frac{\as}{2\pi} \right)^2 \biggl\{2 C_F^2 L_\rho^4    + \biggl[  C_F^2 \left(6 - 16 L_R\right) + C_F C_A \frac{11}{3} - C_F T_R  n_F \frac{4}{3} \biggr]L_\rho^3 \\
&\qquad+ \biggl[ C_F^2 \biggl( \frac{13}{2} - 2\pi^2 - 24L_R + 40L_R^2\biggr) + C_F C_A \biggl(-\frac{169}{36} + \frac{\pi^2}{3} - \frac{44}{3}L_R\biggr)   + C_F T_R n_f \biggl( \frac{11}{9} + \frac{16}{3}L_R\biggr)\biggr]L_\rho^2 \\
& \qquad + \biggl[ C_F^2 \biggl( \frac{9}{4} \minus  2\pi^2 \plus 4\zeta_3 \plus 8(\pi^2 \minus 1)L_R + 12L_R^2  - 32 L_R^3 \biggr) + C_F C_A \biggl(\minus \frac{57}{4} \plus 6\zeta_3\plus \frac{268 \minus 12\pi^2}{9}L_R  \plus  \frac{44}{3}L_R^2 \biggr) \\
&\qquad\quad  + C_F T_R n_F \biggl(5  - \frac{80}{9}L_R -  \frac{16}{3}L_R^2\biggr) + \frac{1}{4}\gamma(R)\biggr]L_\rho\biggr\} \,,
\end{split}
\ee
up to a term constant in $\rho$. Here $\as\equiv\as(Q)$,  $L_\rho \equiv \ln\rho$, $L_R\equiv \ln\tan(R/2)$,  and $\gamma(R)$ is the unknown part of the $\mathcal{O}(\as^2)$ non-cusp anomalous dimensions in \eqs{gammaSin}{gammaSout}. Meanwhile, the out-of-jet contribution is
\be
\begin{split}
\label{vetoglobal}
&\Sigma_{\text{out}}(\Lambda) = 1 - \frac{\alpha_s C_F}{\pi}\biggl[ \frac{\pi^2}{6} + 2\Li_2\!\left(\minus\tan^2\!\frac{R}{2}\!\right)      + 4L_\Lambda L_R + 2L_R^2\biggr] + \left(\frac{\as}{2\pi} \right)^2 \biggl\{ \biggl[32C_F^2 L_R^2  + \frac{44 C_F C_A - 16 C_F T_R n_F}{3}L_R\biggr] L_\Lambda^2 \\
& \qquad + \biggl[C_F^2 \biggl(\frac{8\pi^2}{3} + 32\Li_2\Bigl(\minus\tan^2\frac{R}{2}\Bigr) + 32L_R^2\biggr)L_R + C_F C_A \biggl( \frac{11}{9}\left(\pi^2 + 12\Li_2\Bigl(\minus\tan^2\frac{R}{2}\Bigr)\right)  + \frac{12\pi^2-268}{9}L_R + \frac{44}{3}L_R^2\biggr) \\
& \qquad\quad  + C_F  T_R n_F \biggl( -\frac{4}{9}\left(\pi^2 + 12\Li_2\Bigl(\minus\tan^2\frac{R}{2}\Bigr)\right) + \frac{80}{9}L_R - \frac{16}{3} L_R^2\biggr) - \frac{1}{4}\gamma(R)  \biggr]L_\Lambda \biggr\}\,,
\end{split}
\ee
\end{widetext}
plus a term constant in $\Lambda$. 
Here $L_\Lambda \equiv \ln(2\Lambda/Q)$.
In the following we only study double logs $L_\rho^2$ and $L_\Lambda^2$, so we do not need to know $\gamma(R)$.
Note that $\Sigma_{\text{out}}\to 1$ in the limit $R\to \pi/2$, as it must for hemisphere jets for which there is no ``out'' region.

The non-global term $\Sigma_{\text{NG}}$ in \eq{cumulantfactorization} predicted by the results in \sec{sec:doubleNGLs} and \sec{sec:NGL} is
\be
\label{massvetoNGL}
\Sigma_{\text{NG}}(\rho,\Lambda) = \minus\left(\frac{\as}{2\pi}\right)^2 \! C_F C_A 2f_{\text{OR}}^{\text{alg}}(R,R)  \ln^2\frac{2\Lambda\tan\frac{R}{2}}{Q\rho}\,,
\ee
with $f_{\text{OR}}^{\text{alg}}$ given for the cone and anti-\kt algorithms by \eq{fR}, for the C/A and \kt algorithms by \eq{fCA}.\footnote{Ref.~\cite{Banfi:2010pa} expresses this NGL as a function of $Q\rho/(2\Lambda R^2)$, working in the small $R$ limit. The extra factor of $R$ relative to \eq{massvetoNGL} is due to the imposition in \cite{Banfi:2010pa} of a hard cutoff $Q/2$ on the soft gluon energy in full QCD. Since the angle is cut off by $R$, this  corresponds to  cutting off the virtuality of soft gluons by $k^2 \sim Q^2\tan^2(R/2)$. This choice of cutoff thus corresponds to evaluating the SCET soft function \eq{S2IO} at the scale $\mu = Q\tan(R/2)$. We constructed the global logs in \eq{cumulantfactorization} by running the jet and soft functions to $\mu=Q$, which leads naturally to writing the NGL in the form \eq{massvetoNGL}. Thus the NGLs quoted in \eq{massvetoNGL} and \cite{Banfi:2010pa} are compatible, with the two conventions differing in which terms get grouped into the soft NGLs. Predictions for physical cross sections remain equivalent.}  Soft gluon clustering, present for the C/A and \kt algorithms, also affects the Abelian $C_F^2$ terms in the cross section \cite{Banfi:2005gj,Delenda:2006nf}.  These clustering effects do not come from RG evolution, and can be included as a correction factor similar to $\Sigma_{\text{NG}}$.

\subsection{Comparison to EVENT2}

\begin{figure}[t!]{
\centering
\includegraphics[width=\columnwidth]{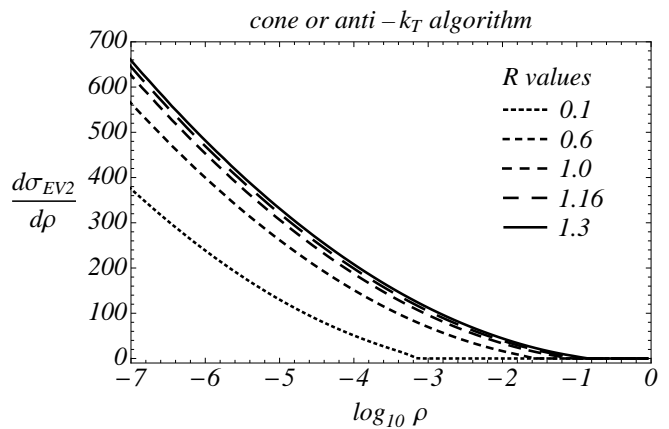}
{ \caption[1]{The $C_F C_A$ terms in the distribution $d\sigma_{\text{EV2}}/d\rho$, using the anti-\kt algorithm.  The coefficient of $\sigma_0(\as/2\pi)^2 C_F C_A$ is plotted.  Five $R$ values are shown, and $\Lambda = 0.01Q$ is used.}
\label{fig:Event2AKTdist}}
}
\end{figure}

For comparison, in \fig{fig:Event2AKTdist} we plot the $C_F C_A$ terms in $d\sigma_{\text{EV2}}/d\rho$ before any terms are subtracted.  This gives the size of all terms, and after subtracting the global logs we can see the size of the non-global terms.

We can test the predictions for the leading NGLs by comparing to the output of EVENT2 \cite{Catani:1996jh,Catani:1996vz}.  EVENT2 can numerically calculate an observable vanishing in the two-jet limit for $e^+e^-$ collisions at $\cO(\as^2)$.  For a given $R$ and $\Lambda$, we can use EVENT2 to find the $\rho$ distribution at $\cO(\as^2)$, $d\sigma_{\text{EV2}}/d\rho$.  By going to the regime $\rho \ll \Lambda \tan R/2$, we can numerically enhance the NGLs relative to the non-logarithmic non-global terms.

EVENT2 calculates a binned distribution in $\rho$, with the cross section in a bin given by
\be
\int_{\rho_{\min}}^{\rho_{\max}} d\rho \, \frac{d\sigma_{\text{EV2}}}{d\rho} = \Sigma_{\text{EV2}} (\rho_{\max}) - \Sigma_{\text{EV2}} (\rho_{\min}) \,,
\ee
with
\be
\Sigma_{\text{EV2}}(\rho) = \int_0^{\rho} d\rho' \, \frac{d\sigma_{\text{EV2}}}{d\rho'} \,.
\ee
In comparing to EVENT2, we will subtract the global and leading NGLs:
\be
\label{Deltasigma}
\Delta \sigma (\rho) \equiv \frac{d\sigma_{\text{EV2}}}{d\rho} - \left( \frac{d\sigma_{\text{global}}}{d\rho} + \frac{d\sigma_{\ln^2}}{d\rho} \right) \,.
\ee
If only single logs remain, then the resulting binned distribution will be constant with value of the single log coefficient.  The non-log terms in the distribution will be a small correction to flatness in the limit $\rho \ll \Lambda \tan R/2$.

\begin{figure*}[h!]{
\centering
\subfigure[]{\scalebox{0.45}{\includegraphics{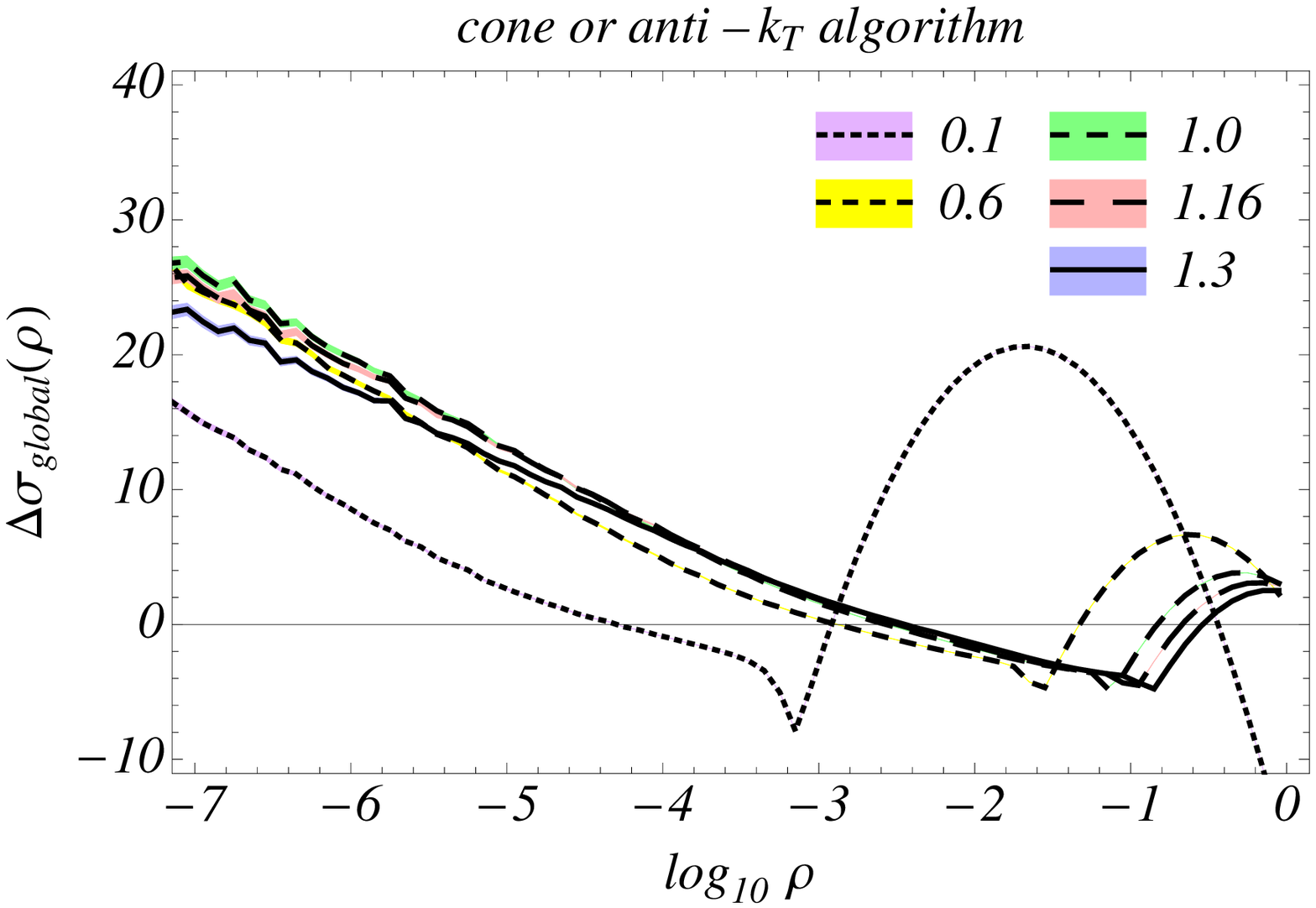}}\label{fig:Event2AKT:global}}
\subfigure[]{\scalebox{0.45}{\includegraphics{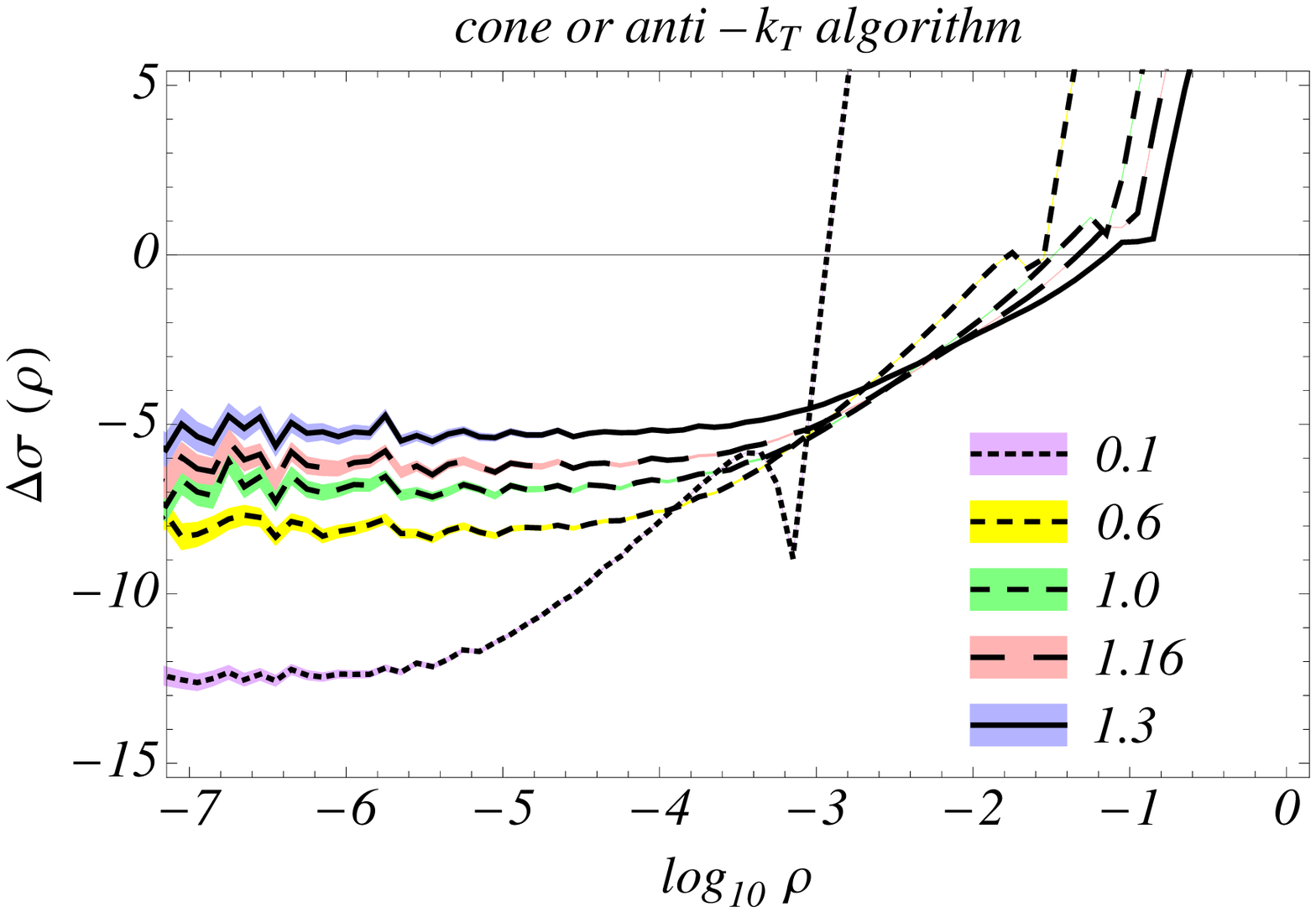}}\label{fig:Event2AKT:NG}}
\vspace{-1em}
{ \caption[1]{The difference between EVENT2 and (a) the global logs and (b) the global and leading NGLs for the $\rho$ distribution, using the anti-\kt algorithm.  The coefficient of $\sigma_0(\as/2\pi)^2 C_F C_A$ in the difference is plotted.  Five $R$ values are shown, and $\Lambda = 0.01Q$ is used.  Each difference becomes flat for small $\rho$, indicating that only single logs in the distribution remain.}
\label{fig:Event2AKT}}
}
\end{figure*}

\begin{figure*}[h!]{
\centering
\subfigure[]{\scalebox{0.45}{\includegraphics{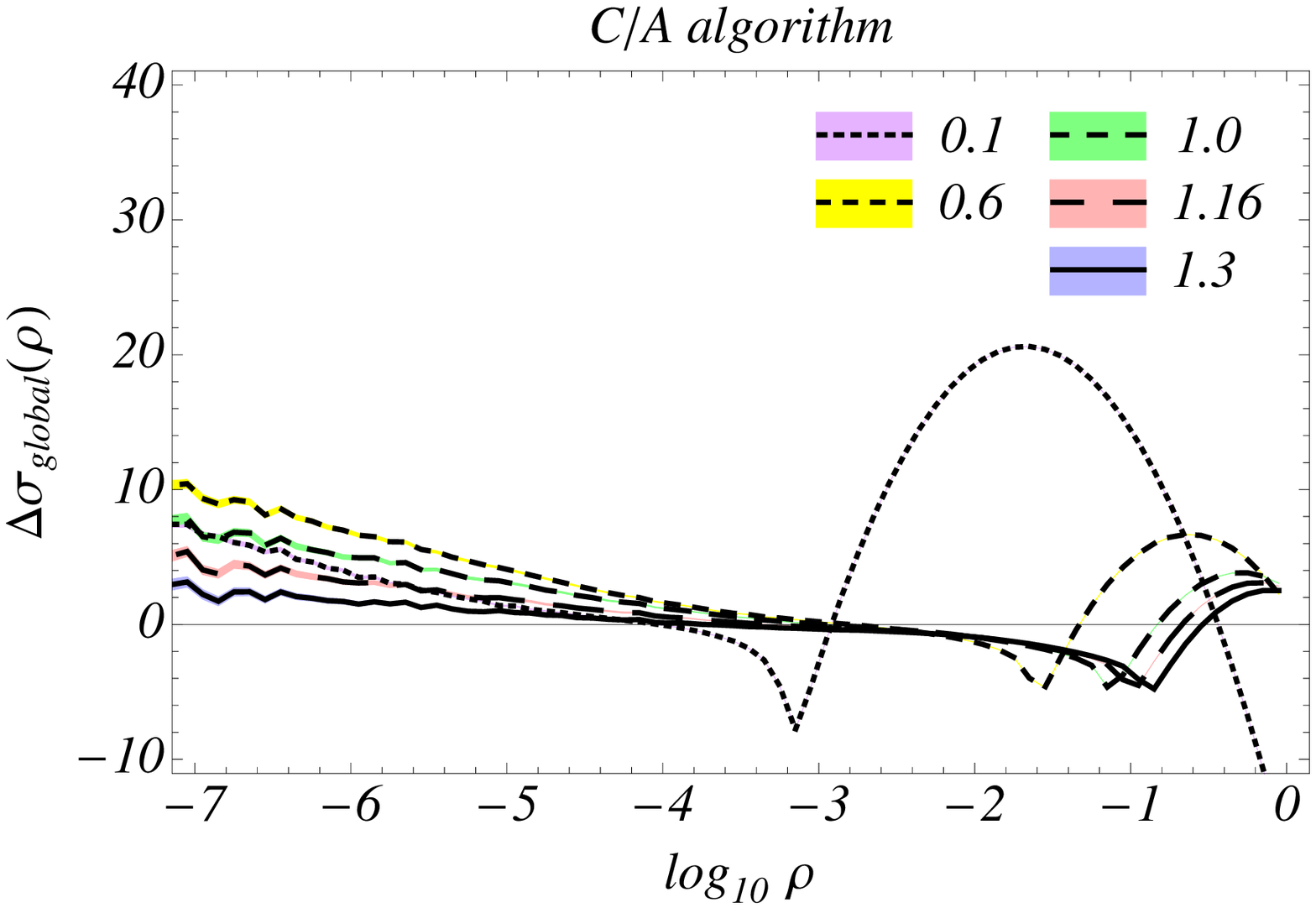}}\label{fig:Event2CA:global}}
\subfigure[]{\scalebox{0.45}{\includegraphics{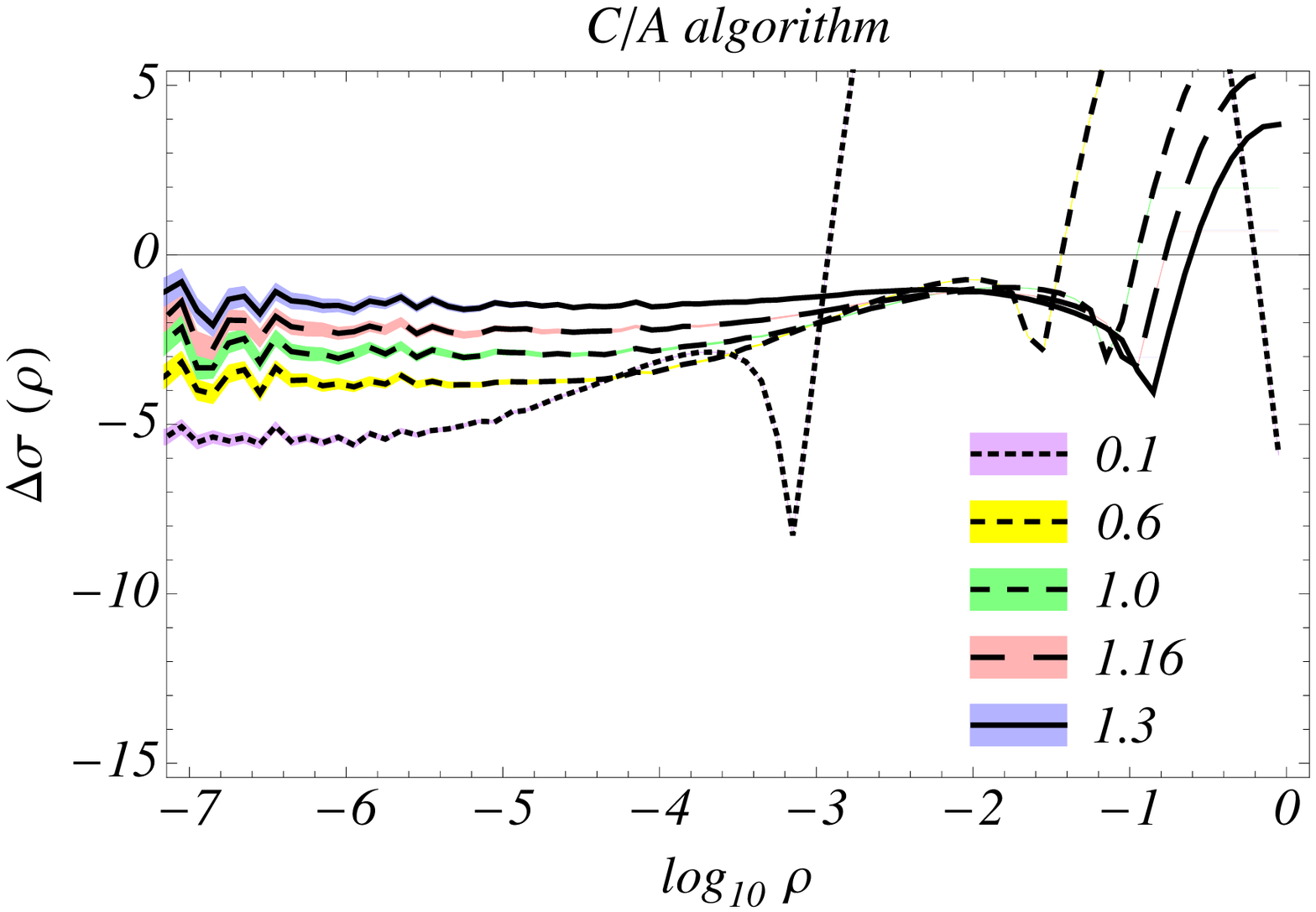}}\label{fig:Event2CA:NG}}
\vspace{-1em}
{ \caption[1]{The difference between EVENT2 and (a) the global logs and (b) the global and leading NGLs for the $\rho$ distribution, using the C/A algorithm.  The coefficient of $\sigma_0(\as/2\pi)^2 C_F C_A$ in the difference is plotted.  Five $R$ values are shown, and $\Lambda = 0.01Q$ is used.  Each difference becomes flat for small $\rho$, indicating that only single logs in the distribution remain.}
\label{fig:Event2CA}}
}
\end{figure*}

\begin{figure*}[h!]{
\centering
\subfigure[]{\scalebox{0.45}{\includegraphics{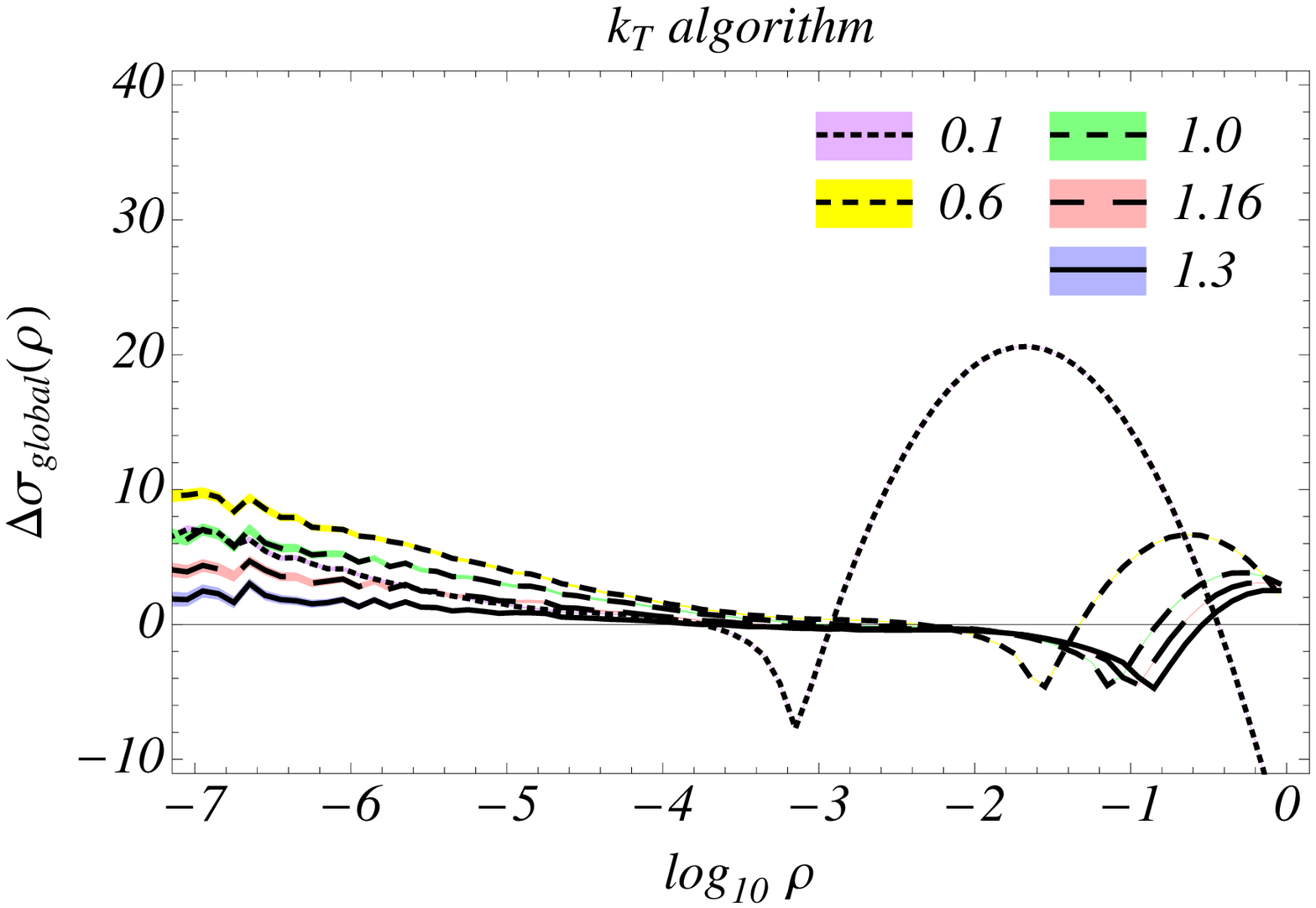}}\label{fig:Event2KT:global}}
\subfigure[]{\scalebox{0.45}{\includegraphics{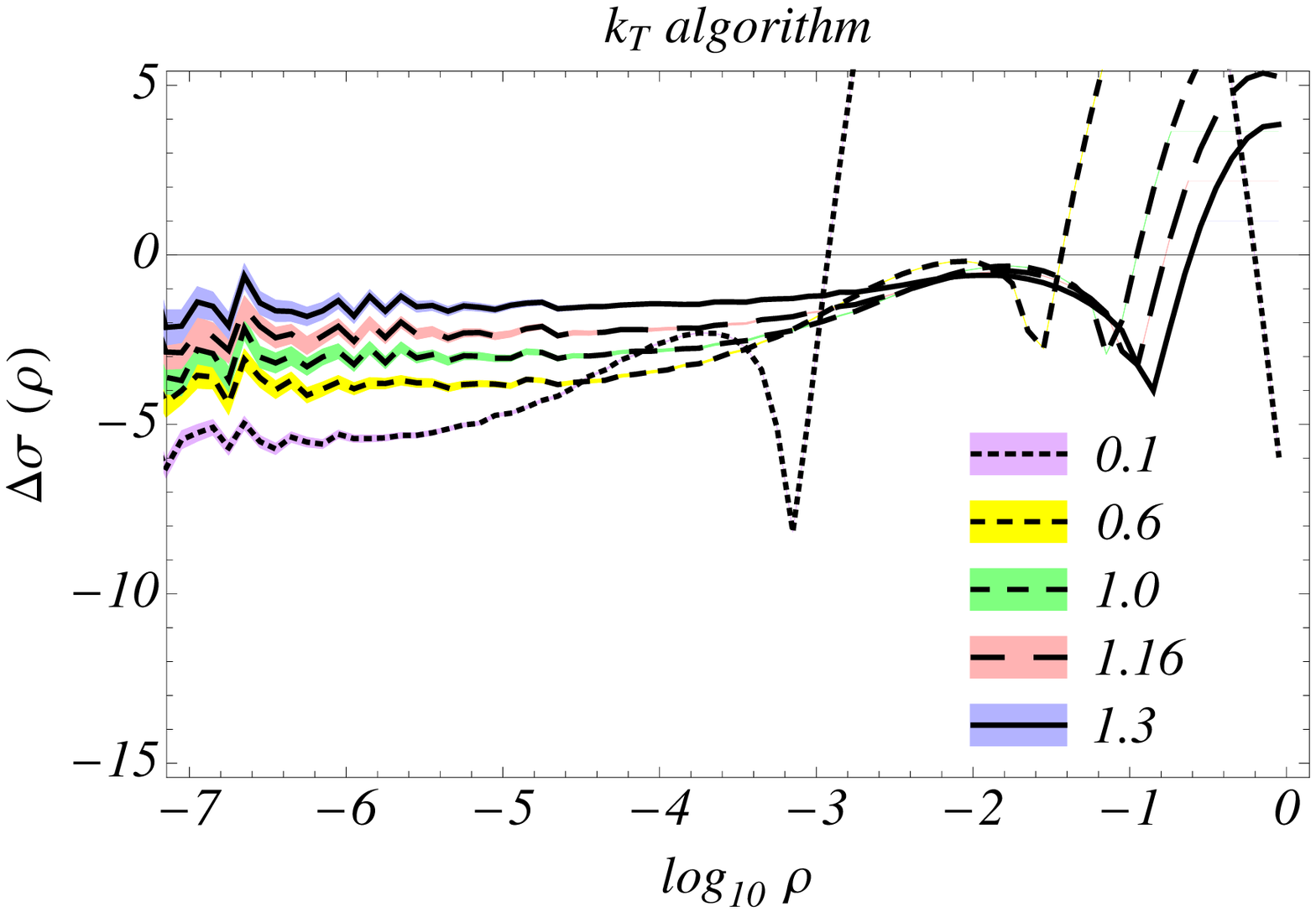}}\label{fig:Event2KT:NG}}
\vspace{-1em}
{ \caption[1]{The difference between EVENT2 and (a) the global logs and (b) the global and leading NGLs for the $\rho$ distribution, using the \kt algorithm.  The coefficient of $\sigma_0(\as/2\pi)^2 C_F C_A$ in the difference is plotted.  Five $R$ values are shown, and $\Lambda = 0.01Q$ is used.  Each difference becomes flat for small $\rho$, indicating that only single logs in the distribution remain.}
\label{fig:Event2KT}}
}
\end{figure*}

For each algorithm, we calculate $\Delta \sigma(\rho)$ for $\Lambda = 0.01Q$ and five values of $R$: $R = \{ 0.1, 0.6, 1.0, 1.16, 1.3 \}$.  $R = 1.16$ corresponds to the jet radius used in the EVENT2 studies in~\cite{Kelley:2011tj}.  For each algorithm and $R$ value, we plot in \fig{fig:Event2AKT}a, \fig{fig:Event2CA}a and \fig{fig:Event2KT}a the $C_F C_A$ terms in $\Delta\sigma_{\text{global}}$ with only the global logs in \eq{Deltasigma} subtracted out, plotted in units of $\sigma_0 (\as/2\pi)^2 C_F C_A$.  The resulting distributions grow linearly for small $\rho$, indicating the presence of a remaining double log.
In \fig{fig:Event2AKT}b, \fig{fig:Event2CA}b and \fig{fig:Event2KT}b, we plot the $C_F C_A$ terms in $\Delta \sigma(\rho)$ after also subtracting out the double NGL in \eq{Deltasigma}. 
The distribution is convincingly flat in the small $\rho$ regime for all three algorithms.  This means that the double logs of $\rho$ have been successfully removed, confirming the calculations of the NGLs performed in \sec{doubleNGLs}.

In \fig{fig:Event2AKT}, we plot the $C_F C_A$ terms in both $\Delta \sigma_{\text{global}} (\rho)$ and $\Delta \sigma(\rho)$, where $\Delta \sigma_{\text{global}}$ only has the global logs removed (and not the leading NGLs).  We can clearly see the double log dependence in $\Delta \sigma_{\text{global}}$, and that it is entirely removed within statistical uncertainties in $\Delta \sigma$.  Note that as $R$ decreases, if we want the non-global double log to be numerically large then the range of $\rho$ must move to smaller values.

In \figs{fig:Event2CA}{fig:Event2KT}, we plot the $C_F C_A$ terms in $\Delta \sigma(\rho)$ for the C/A and \kt algorithms.  As with anti-\kt, we can see that the double log dependence has been removed within statistical uncertainties.  Additionally, we confirm the prediction that the leading NGL is the same for the C/A and \kt algorithms in the limit $\rho \ll \Lambda$.

\subsection{Remarks on Soft Factorization}

KSZ proposed that the soft function $S(k_L,k_R,\Lambda)$ in \eq{conesoftdef}, at least in the regime $\Lambda< k_{L,R}\ll QR\ll Q$, factorizes to all orders in $\as$ into an in-jet and out-of-jet piece,
\be
\label{KSZansatz}
S(k_L,k_R,\Lambda) = S_{\text{in}}(k_L,k_R)S_{\text{out}}(\Lambda)\,,
\ee
without any additional factor like $S_{\text{NG}}$ in our \eq{factorizedsoft} containing logs of $\Lambda/k_{L,R}$.
KSZ recognized that $S_{\text{in}}(k_L,k_R)$ can contain logs of $k_L/k_R$ 
and does not factorize na\"{i}vely, but went on to claim that the out-of-jet piece factors off completely. Their proof relied on the absence of any logs of $\Lambda/k_{L,R}$ in $S(k_L,k_R,\Lambda)$. We have shown above that such NGLs do in fact arise at $\cO(\as^2)$ for any value of the ratio $\Lambda/k_{L,R}$ and  that therefore \eq{KSZansatz} does not hold. Instead $S$ takes the form we derived in \eq{factorizedsoft}.

KSZ provided evidence for their claim by comparing their prediction, assuming NGLs of $\Lambda/k_{L,R}$ are absent,  to the $\cO(\as^2)$ QCD prediction from EVENT2, using the C/A algorithm with $R=1.16$ and $\Lambda = 0.01Q$.  From our calculation and EVENT2 analysis, we observe that the C/A NGL coefficient for this choice of $R$ and $\Lambda$ is too small to be resolved in the corresponding plot of \cite{Kelley:2011tj} (Fig.~1 in v1 or Fig.~7 in v2), which plots the distribution down to $\rho\sim 10^{-4}$.  Our calculation of $f_{\text{OR}}^{\text{C/A}}$ in \eq{fCA} and comparison to EVENT2 shown in \fig{fig:Event2CA} demonstrate that the NGL $2\as^2 C_F C_A f_{\text{OR}}^{\text{C/A}}(R) \ln^2(2\Lambda\tan\frac{R}{2}/Q\rho)$ is clearly present (which, as one realizes from our calculations above, is also implicit in the results of \cite{Banfi:2010pa}). Furthermore, based on the results for NGLs in hemisphere soft functions in \cite{Kelley:2011ng,Hornig:2011iu}, we expect single NGLs of $\Lambda/k_{L,R}$ and non-global non-logarithmic functions of $\Lambda/k_{L,R}$ to arise in $S(k_L,k_R,\Lambda)$ and also violate the KSZ ansatz \eq{KSZansatz}.

\section{Conclusions}
\label{sec:conclusions}

We have derived the leading NGLs of ratios of jet masses $m_{1,2}$ and an energy veto $\Lambda$ on additional jets in dijet cross sections $\sigma(m_1^2,m_2^2,\Lambda)$ using  several different jet algorithms. We confirm earlier qualitative findings about the effects of jet sizes $R$ and clustering in recombination algorithms on the size of NGLs involving a jet veto parameter, and in addition provide the full algorithmic and $R$ dependence for NGLs involving both jet masses and vetoes for the first time. Ours is also the first explicit calculation in the framework of effective field theory of the leading $\cO(\as^2)$ NGLs in a soft function appearing in the factorization theorem for an exclusive jet cross section involving a jet veto.

We confirm the insight of \cite{Appleby:2002ke} that the action of clustering by recombination algorithms reduces the size of NGLs in general. The larger the phase space in which soft gluons can be recombined, the smaller the NGLs.  This is also borne out by the $R$ dependence of the in-out NGLs and in-in NGLs.

Our calculation makes clear that NGLs arise from soft gluons anywhere in the separated regions of phase space into which they are allowed to go, and not only from those  splitting right along the boundary. Although the latter give the largest numerical contribution, as seen in the behavior of \eq{fOR} as $\eta_1-\eta_2\to 0$, the enhancement is not parametrically large, and gluons with significant angular separation still give non-negligible contributions to NGL. Thus observables sensitive to soft gluons in any separated regions of phase space probed with different soft scales are prone to NGLs.  From the EFT point of view, sensitivity of soft modes to any disparate soft scales generates non-global dependence on the ratio of those scales in the soft function. 

Going beyond the explicit calculations of the leading NGLs of jet masses and vetoes, we uncovered strong relations among the coefficients of the in-in and in-out NGLs and the contributions to NGLs from diagrams with gluons in the same region of phase space (both gluons in the same jet or both outside the jets). We did this by extending our insight in \cite{Hornig:2011iu} that in EFT, NGLs are built out of contributions from regions of phase space where two gluons enter separated regions or the same region, each contributing a log of the factorization scale $\mu$ over a single scale---$k_{L,R}$ if both gluons enter the same jet, $\Lambda$ if both lie outside the jets, and intermediate scales $\sqrt{k_{L,R}\Lambda}$ or $\sqrt{k_L k_R}$ if the gluons enter separated regions (cf. \eq{NGLterms}). IR safety of the soft function and RG invariance require the coefficients of these logs all to be related so that they add up to the $\mu$-independent NGLs of $k_L/k_R$ and $\Lambda/k_{L,R}$. By considering the possible dependence of these coefficients on separate jet radii $R_{L,R}$, we derived new relations among the coefficients of all the different contributions to NGLs.

The above lessons about the properties of and methods to calculate NGLs are directly applicable to exclusive jet cross sections in hadron collisions such as at the LHC. Gaining control of and resumming NGLs or using methods that minimize their impact on jet cross sections will be essential to achieving precise theoretical predictions.

In the quest to resum NGLs using an EFT framework, a physical picture is important to recognize non-global observables and understand the implication for factorization theorems.  While some elements of the physical picture drawn above have previously been noticed, they are often not appreciated sufficiently so as to make obvious the structure of NGLs and when they will appear in cross sections.  We have studied in detail the NGLs that appear in measurements of jet masses in exclusive jet cross sections with a jet veto.  We hope that this work has made clear the source of NGLs, providing both better intuition for non-global observables and a quantitative EFT-based approach to study them.

\begin{acknowledgments}

We would like to thank the organizers of the SCET 2011 Workshop hosted by Carnegie Mellon University and the University of Pittsburgh where some results of this work were first derived and presented, and the Institute for Nuclear Theory and the organizers of the INT program on ``Frontiers of QCD'' where this work was completed. We are indebted to Iain Stewart for collaboration on related work and helpful comments on a draft of this paper. We thank Randall Kelley and Matt Schwartz for comments on the results of \cite{Kelley:2011tj}. We are grateful to Mrinal Dasgupta, Gavin Salam and Frank Tackmann for helpful feedback, and of course, we thank Zoltan Ligeti. This work is supported in part by the Offices of Nuclear and High Energy Physics of the U.S. Department of Energy under Contracts DE-FG02-96ER40956, DE-FG02-94ER40818, and DE-AC02-05CH11231.  The work of JW was supported in part by a LHC Theory Initiative Postdoctoral Fellowship, under the National Science Foundation grant PHY-0705682.

\end{acknowledgments}

\appendix

\section{Global Logs from RGE}
\label{appx:global}

The parts of the cross section \eq{cumulant} predicted by RG evolution are given by (cf. \cite{Becher:2006nr,Ellis:2009wj,Ellis:2010rw})
\be
\label{resummedcumulantin}
\begin{split}
&\Sigma_{\text{in}}(\rho) = \sigma_0 H(\mu_H) e^{K_H(\mu,\mu_H)+2K_J(\mu,\mu_J)+ 2K_S^{\text{in}}(\mu,\mu_S)}  \\
&\quad \times \left(\!\frac{\mu_H}{Q}\!\right)^{\!\!\omega_H(\mu,\mu_H)}\! \! \left(\!\frac{\mu_J^2}{Q^2\rho}\!\right)^{\!\!2\omega_J(\mu,\mu_J)}\! \! \left(\!\frac{\mu_S \tan\frac{R}{2}}{Q\rho}\!\right)^{\! \!2\omega_S^{\text{in}}(\mu,\mu_S)\minus\Omega}\\
&\qquad\qquad \times \tilde J^2\left(\ln\frac{\mu_J^2}{Q\mu_S\tan\frac{R}{2}} + \partial_\Omega \right) \widetilde S^2_{\text{in}}(\partial_\Omega) \\
&\qquad\qquad \times \left[\left(\frac{\mu_S\tan\frac{R}{2}}{Q\rho}\right)^{\!\Omega}\! \frac{e^{\gamma_E \Omega}}{\Gamma(1-\Omega)}\right]\,
\end{split}
\ee
and
\be
\label{resummedcumulantout}
\begin{split}
\Sigma_{\text{out}}(\rho) =  S_{\text{out}}(\mu_\Lambda)\left(\frac{\mu_\Lambda}{2\Lambda}\right)^{\omega_S^{\text{out}}(\mu,\mu_\Lambda)}e^{K_S^{\text{out}}(\mu,\mu_\Lambda)}\,,
\end{split}
\ee
where in \eqs{resummedcumulantin}{resummedcumulantout}, $\Omega = 2\omega_J (\mu,\mu_J)+ 2\omega_S^{\text{in}}(\mu,\mu_S)$, and each factor $K_F \equiv K_F(\mu,\mu_F),\omega_F \equiv \omega_F(\mu,\mu_F)$ is given by the cusp and non-cusp parts of the anomalous dimension of $F$ by 
\begin{subequations}
\label{Komegadefs}
\begin{align}
\label{Kint}
K_F(\mu,\mu_0)  &=  \int_{\mu_0}^\mu \frac{d\mu'}{\mu'}\Bigl(2\Gamma_{F}[\alpha_s(\mu')]\ln\frac{\mu'}{\mu_0}+ \gamma_F[\alpha_s(\mu')]\Bigr), \\
\label{omegaint}
\omega_F(\mu,\mu_0)  &= \frac{2}{j_F}\int_{\mu_0}^\mu\frac{d\mu'}{\mu'}\Gamma_{F}[\alpha_s(\mu')]\,,
\end{align}
\end{subequations}
which are given to $\cO(\as^2)$ by
\begin{subequations}
\label{Komegaexpansions}
\begin{align}
\label{Kdef}
K_F(\mu,\mu_F) &=  \frac{\as(\mu)}{4\pi} \left(\Gamma_F^0 \ln^2\frac{\mu}{\mu_F}+ \gamma_F^0 \ln\frac{\mu}{\mu_F}\right) \nn \\
&+ \left(\frac{\as(\mu)}{4\pi} \right)^2 \biggl[ \frac{2}{3} \Gamma_F^0 \beta_0 \ln^3\frac{\mu}{\mu_F} \\
&\qquad + (\gamma_F^0 \beta_0 + \Gamma_F^1) \ln^2\frac{\mu}{\mu_F} + \gamma_F^1\ln\frac{\mu}{\mu_F}\biggr]  \nn \\
\label{omegadef}
\omega_F(\mu,\mu_F) &=  \frac{2}{j_F}\biggl\{ \frac{\as(\mu)}{4\pi} \Gamma_F^0 \ln\frac{\mu}{\mu_F}  \\
&+ \left(\frac{\as(\mu)}{4\pi} \right)^2 \biggl[ \Gamma_F^0 \beta_0 \ln^2\frac{\mu}{\mu_F} + \Gamma_F^1 \ln\frac{\mu}{\mu_F}\biggr] \biggr\} \,, \nn
\end{align}
\end{subequations}
where the constant $j_F$ in \eqs{omegaint}{omegadef} is $2$ for $F=J$ and 1 otherwise.

The anomalous dimensions $\Gamma_F,\gamma_F$ appearing in \eqs{Komegadefs}{Komegaexpansions} are given by
\be
\Gamma_F(\alpha_s) = \sum_{n=0}^\infty \left(\frac{\as}{4\pi}\right)^{n+1}\Gamma_F^n \ , \ \gamma_F(\alpha_s) = \sum_{n=0}^\infty \left(\frac{\as}{4\pi}\right)^{n+1}\gamma_F^n \,,
\ee
where the $\Gamma_F$ pieces for each function $F$ are proportional to the cusp anomalous dimension,
\be
\Gamma_J = 2 C_F \Gamma_{\text{cusp}} \ , \ \Gamma_S^{\text{in}} = - C_F \Gamma_{\text{cusp}} \,,
\ee
with
\be
\Gamma_{\text{cusp}}^0 = 4 \ , \ \Gamma_{\text{cusp}}^1 = 4 C_A\left(\frac{67}{9} - \frac{\pi^2}{3}\right) -  \frac{40}{9}n_F \, ,
\ee
and the non-cusp anomalous dimensions are given for the inclusive jet function by
\begin{subequations}
\begin{align}
\gamma_J^0 &= 6 C_F \\
\gamma_J^1 &= C_F^2( 3 \minus 4\pi^2 \plus 48\zeta_3) + C_F C_A\left(\frac{1769}{27} \plus \frac{22\pi^2}{9} \minus 80\zeta_3\right) \nn \\
&\qquad  - C_F T_R n_F\left(\frac{484}{27} + \frac{8\pi^2}{9}\right)
\end{align}
\end{subequations}
for the in-cone soft function by
\begin{subequations}
\label{gammaSin}
\begin{align}
\gamma_S^0 &=0 \\
\begin{split}
\gamma_S^1 &=  C_F C_A\left(-\frac{808}{27} + \frac{11\pi^2}{9} +28\zeta_3\right) \\
&\qquad + C_F  T_R n_F\left(\frac{224}{27} - \frac{4\pi^2}{9}\right) - \frac{1}{2}\gamma(R) \,,
\end{split}
\end{align}
\end{subequations}
and for the out-of-cone soft function by
\be
\label{gammaSout}
\Gamma_{\text{out}} = 0 \ , \ \gamma_{\text{out}} = 2 C_F\Gamma_{\text{cusp}} \ln\tan^2\frac{R}{2} + \gamma(R) \,.
\ee
The anomalous dimensions \eqs{gammaSin}{gammaSout} have not yet been calculated explicitly  beyond $\cO(\as)$ for jets with non-hemisphere radii $R$, but can be deduced from the known hard and inclusive jet anomalous dimensions to $\cO(\as^3)$ and consistency of RG evolution. 
The relation \eq{gammaSout} between $\gamma_{\text{out}}$ and the universal cusp anomalous dimension was proposed in \cite{Ellis:2009wj,Ellis:2010rw} based on consistency of the RG evolution in the factorization theorem \eq{factorization}. The term $\gamma(R)$ is not yet known, but it must vanish at $R=\pi/2$  and should be non-singular as $R\to 0$. It must also cancel in the sum $2\gamma_S^1 + \gamma_{\text{out}}$. 

The hard anomalous dimensions are constrained by the requirement of consistency of RG running to be $\Gamma_H = - 2\Gamma_J - 2\Gamma_S^{\text{in}}$ and $\gamma_H = -2\gamma_J - 2\gamma_S^{\text{in}} - \gamma_S^{\text{out}}$. 
We also use the beta function coefficient $\beta_0 = (11 C_A - 2n_F)/3$.

The fixed-order hard, jet, and soft functions in \eqs{resummedcumulantin}{resummedcumulantout} are all given to $\cO(\as^2)$ by 
\begin{align}
\label{fixedorderform}
&F(\mu_F) = 1 + \frac{\as(\mu_F)}{4\pi}\left(\Gamma_F^0 \ln^2\frac{\mu_F}{Q_F} + \gamma_F^0 \ln\frac{\mu_F}{Q_F} + c_F^1\right) \nn \\
& + \left(\!\frac{\as(\mu_F)}{4\pi}\!\right)^2 \biggl[ \frac{1}{2} \left(\Gamma_F^0\right)^2 \ln^4\frac{\mu_F}{Q_F} + \left(\Gamma_F^0 \gamma_F^0 + \frac{2}{3}\Gamma_F^0\beta_0\right)\ln^3\frac{\mu_F}{Q_F} \nn \\
&+ \left(\frac{1}{2}(\gamma_F^0)^2 + \gamma_F^0\beta_0 + \Gamma_F^1 + c_F^1\Gamma_F^0 \right)\ln^2\frac{\mu_F}{Q_F}  \nn\\
&+ (\gamma_F^1 + c_F^1 \gamma_F^0 + 2c_F^1 \beta_0)\ln\frac{\mu_F}{Q_F} + c_F^2\biggr] \,.
\end{align}
For the jet and in-cone soft functions $\tilde J,\tilde S_{\text{in}}$ in \eq{resummedcumulantin}, each log in \eq{fixedorderform} should be replaced by the differential operator appearing in the arguments given in \eq{resummedcumulantin}.
The scale $Q_F$ appearing in the logs for each function $F=H,J,S_{\text{in}},S_{\text{out}}$ is
\be
Q_H = Q \ , \ Q_J = Q\sqrt{\rho} \ , \ Q_S^{\text{in}} = \frac{Q\rho}{\tan\frac{R}{2}} \ , \ Q_S^{\text{out}} = 2\Lambda\,.
\ee
Thus it is most natural to evaluate each function $F(\mu_F)$ in \eqs{resummedcumulantin}{resummedcumulantout} at the canonical scale $\mu_F = Q_F$, but the expressions  \eqs{resummedcumulantin}{resummedcumulantout} are invariant under different choices of $\mu_F$.

\begin{figure*}[th!]{
    \includegraphics[width=0.95\textwidth]{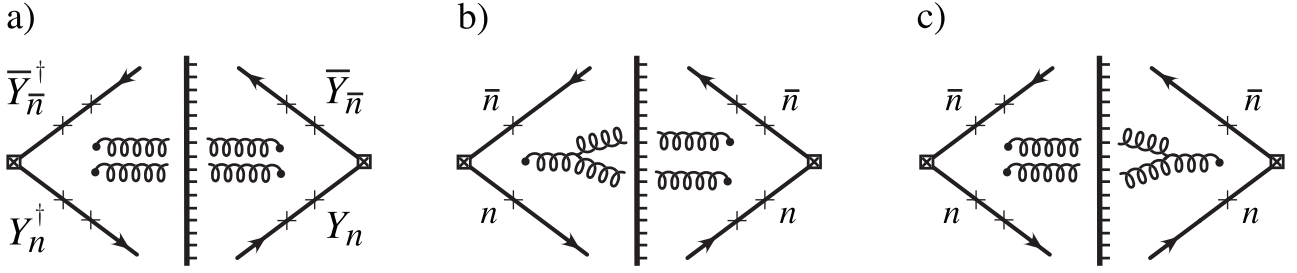} { \caption[1]{$\cO(\as^2)$
        real emission diagrams contributing to leading NGLs.  The endpoints of the gluons can be
        attached to the points on the Wilson lines labeled by a `x' in any
        order.  Figure (a) gives the $\cI$ diagrams, (b) and (c) give the $\cT$
        diagrams.}
  \label{fig:diagrams}} }
\end{figure*}

The constants $c_F^1$ in the $\cO(\as)$ fixed-order functions $F$ are given by
\begin{subequations}
\begin{align}
c_H^1 &= -C_F\left(16-\frac{7\pi^2}{3}\right) \\
c_J^1 &= C_F\left( 7 - \pi^2\right) + \frac{\Gamma_J^0}{4}\frac{\pi^2}{6} \\
c_{S_{\text{in}}}^1 &= C_F\frac{\pi^2}{6} + \Gamma_S^0\frac{\pi^2}{6} \\
c_{S_{\text{out}}}^1 &= -C_F\left[ 2\ln^2\tan^2\frac{R}{2} +  \frac{2\pi^2}{3} +8 \Li_2\!\left(\minus\tan^2\frac{R}{2}\right)\right]\,.
\end{align}
\end{subequations}
The one-loop in- and out-of-cone soft function constants were first derived in \cite{Ellis:2010rw}. The two-loop constants $c_F^2$ in \eq{fixedorderform} are known for the hard \cite{Becher:2006mr} and jet \cite{Becher:2006qw} functions, but not yet for the in- and out-of-cone soft functions. We do not need the constants $c_F^2$ in this paper since we only study the logarithmic behavior of $\Sigma(\rho,\Lambda)$ at $\cO(\as^2)$.

\section{Leading Non-Global Contributions to Soft Function at $\cO(\as^2)$}
\label{appx:NGL}

The ``In-Out'' or ``In-In'' soft functions in \sec{sec:doubleNGLs} and \sec{sec:NGL} are defined by
\be
\label{SNG}
\begin{split}
&S_{\text{NG}}^{\text{OL,OR,LR}} (k_{L,R},\Lambda) = \sum_{j} \int\frac{d^D k_1}{(2\pi)^D}\frac{d^D k_2}{(2\pi)^D} \\
&\quad \times  \cA_j(k_1,k_2)  \cM_{k_1,k_2}^{[OL,OR,LR]} (k_{L,R},\Lambda) 
\cC(k_1)\cC(k_2)
\end{split}
\ee
where  $j = \{\cI,\cT,\cG,\cH,\cQ\}$  sums over different $\cO(\as^2)$ cut Feynman diagram topologies (independent emission, three-gluon vertex, gluon bubble, ghost bubble, and quark bubble, respectively) with two real gluons in the final state. The $\cI,\cT$ diagrams are illustrated in \fig{fig:diagrams}.  The cut propagators give factors $\cC(k_{1,2}) = 2\pi\delta(k_{1,2}^2) \theta(k_{1,2}^0)$, where $k_i^0 = (n\mcdot k_i + \bn\mcdot k_i)/2$. The functions $\cM^{\text{[OL,OR,LR]}}$ impose the measurements on the two final-state gluons,  giving the contributions of the possible ways that two gluons can go into separate regions.
\begin{subequations}
\label{measurements}
\begin{align}
\cM_{k_1,k_2}^{[OL]}(k_{L,R},\Lambda) &= \delta(k_L \minus \Theta_L n\mcdot k_1) \delta(k_R) \delta(\Lambda \minus \Theta_{\text{out}} k_2^0) \nn \\ 
&\quad + (1\!\leftrightarrow\! 2) 
\\
\cM_{k_1,k_2}^{[OR]}(k_{L,R},\Lambda) &= \delta(k_L) \delta(k_R \minus \Theta_R \bn\mcdot k_2) \delta(\Lambda \minus \Theta_{\text{out}} k_2^0) \nn \\
&\quad + (1\!\leftrightarrow\! 2) 
\\
\cM_{k_1,k_2}^{[LR]}(k_{L,R},\Lambda) &= \delta(k_L \minus \Theta_L n\mcdot k_1) \delta(k_R \minus \Theta_R  \bn\mcdot k_2)  \delta(\Lambda) \nn \\
&\quad \plus (1\!\leftrightarrow\! 2)\,,
\end{align}
\end{subequations}
where $\Theta_{L,R}$ are the phase space constraints restricting gluon 1 or 2 to be in the left or right jet, and $\Theta_{\text{out}}$ restricts gluon 1 or 2 to be outside both jets. Their form depends on the algorithm used to find the jets.

Expressions for all the matrix elements $\cA_j$ are given in Appendix B of Ref.~\cite{Hornig:2011iu}. In general covariant gauge, the leading NGL comes only from the sum of $\cI$ and $\cT$ diagrams proportional to the color factor $C_F C_A$. This contribution to the total amplitude is
\be
\label{NGLamplitude}
\begin{split}
(\cA_{\cI_{C_FC_A}} + \cA_{\cT})^{\text{leading}} = 4 &g^4 C_F C_A  \mu^{4\epsilon} \frac{k_1^\perp\mcdot k_2^\perp}{k_1\mcdot k_2} \\
&\times \frac{1}{n\mcdot k_1 \, \bn\mcdot k_1 \,  n\mcdot k_2 \, \bn\mcdot k_2} \,,
\end{split}
\ee
 in $D = 4-2\epsilon$ dimensions. Since the matrix elements are symmetric in $1\leftrightarrow 2$, the symmetrized terms in \eq{measurements} just give factors of 2.

The leading contribution to the non-global parts of the soft function $S_{\text{NG}}^{\text{OR}}$ in \eq{SNG} is then given, after integrating over $k_1^+,k_2^-$ and $k_{1,2}^\perp$ and converting to the $\MSbar$ scheme, by
\be
\label{SORlightcone}
\begin{split}
&S_{\text{NG}}^{\text{OR}}(k_{L,R},\Lambda) =  \frac{\as^2 C_F C_A}{\pi^2}\frac{(\mu^2 e^{\gamma_E})^{2\epsilon}}{\Gamma(1-\epsilon)^2}  \delta(k_L) \\
&\times \int_0^\infty dk_1^- (k_R k_1^-)^{-1-\epsilon} \int_0^\infty dk_2^+ (k_2^+)^{-1-\epsilon} (2\Lambda - k_2^+)^{-1-\epsilon} \\
&\times \frac{\Gamma(1-\epsilon)}{\sqrt\pi\Gamma\left(\frac{1}{2}-\epsilon\right)}\int_0^\pi d\phi \frac{z \sin^{-2\epsilon}\phi}{1 + z^2 - 2z \cos\phi} \Theta_R \Theta_{\text{out}}\,,
\end{split}
\ee
where $z \equiv \sqrt{k_1^- k_2^+/[k_R(2\Lambda-k_2^+)]}$. A similar formula holds for $S_{\text{NG}}^{\text{OL}}$. 
Meanwhile, the leading contribution to $S_{\text{NG}}^{\text{LR}}$ takes the form
\be
\label{SLRlightcone}
\begin{split}
&S_{\text{NG}}^{\text{LR}}(k_{L,R},\Lambda) =  \frac{\as^2 C_F C_A}{\pi^2}\frac{(\mu^2 e^{\gamma_E})^{2\epsilon}}{\Gamma(1-\epsilon)^2}  \delta(\Lambda) \\
&\times \int_0^\infty dk_1^- (k_R k_1^-)^{-1-\epsilon} \int_0^\infty dk_2^+  (k_L k_2^+)^{-1-\epsilon}  \\
&\times \frac{\Gamma(1-\epsilon)}{\sqrt\pi\Gamma\left(\frac{1}{2}-\epsilon\right)} \int_0^\pi d\phi \frac{w \sin^{-2\epsilon}\phi}{1 + w^2 - 2w \cos\phi} \Theta_R \Theta_L\,,
\end{split}
\ee
where $w \equiv \sqrt{k_1^- k_2^+/(k_R k_L)}$.

For the cone or anti-\kt algorithms, the theta functions $\Theta_{R,L,\text{out}}$ take the form
\begin{subequations}
\begin{align}
\Theta_R & \!=\! \theta\Bigl(\frac{k_R}{k_1^-}\!\! <\!  \tan^2\!\frac{R_R}{2}\Bigr)  , \, \Theta_L \!=\! \theta \Bigl(\frac{k_L}{k_2^+} \! \! < \!  \tan^2\!\frac{R_L}{2}\Bigl)  \\
\Theta_{\text{out}} &=  \theta\Bigl(\frac{k_2^+}{2\Lambda \minus k_2^+} \!  > \! \tan^2\frac{R_R}{2} \Bigr)\theta\Bigl(\frac{2\Lambda \minus k_2^+}{k_2^+}\! > \! \tan^2\frac{R_L}{2} \Bigr) .
\end{align}
\end{subequations}
Then it becomes natural to rescale variables in \eqs{SORlightcone}{SLRlightcone} by $k_1^- = k_R x/\tan^2\frac{R_R}{2}$, and $k_2^+ = 2\Lambda y$ or $k_2^+ = k_L y/\tan^2\frac{R_L}{2}$. We group the factors of $R_{L,R}$ and $2\Lambda$ with the $\epsilon$-dependent prefactors in \eqs{SORlightcone}{SLRlightcone} and expand the remaining integrand to leading order in $\epsilon$, as the subleading terms do not contribute to the leading NGL. This results in the particular group of terms appearing in  $S_{\text{NG}}^{\text{OR,LR}}$ given in \eqs{S2IO}{S2LR}, which we have generalized to other algorithms. We also changed variables from the remaining light-cone momenta in \eqs{SORlightcone}{SLRlightcone} to the rapidities of gluons 1 and 2 relative to jet axis 1 to express the coefficients $f_{\text{OR,OL,LR}}$ in the form given in  \eq{fOR}.

\bibliography{NGL}

\end{document}